\documentclass[pre,showkeys,preprintnumbers,amsmath,amssymb,superscriptaddress,twocolumn]{revtex4}
\usepackage{graphicx}
\usepackage{amsmath}
\usepackage{amsfonts}
\usepackage{amssymb}
\usepackage{bm}
\usepackage{color}

\def\n{\bm{n}}
\def\s{\bm{s}}
\def\x{\bm{x}}

\def\I{{\bf I}}
\def\KK{{\bf K}}
\def\MM{{\bf M}}
\def\K{{\mathcal K}}

\def\M{{\mathcal M}}

\def\R{{\mathbb R}}

\def\C{{\mathbb C}}

\def\G{\hat{G}_{\rm f}}
\def\j{\hat{j}_{\rm f}}

\def\pa{{\partial\Omega}}
\def\ve{\varepsilon}

\def\erfc{\mathrm{erfc}}
\def\erfcx{\mathrm{erfcx}}

\def\nmax{n_{\rm max}}

\begin{document}

\title{Spectral theory of imperfect diffusion-controlled reactions \\ on heterogeneous catalytic surfaces}

\author{Denis S. Grebenkov}
  \email{denis.grebenkov@polytechnique.edu}
\affiliation{Laboratoire de Physique de la Mati\`ere Condens\'ee (UMR 7643), CNRS -- Ecole Polytechnique, IP Paris, 91128 Palaiseau, France}

\date{\today}

\begin{abstract}
We propose a general theoretical description of chemical reactions
occurring on a catalytic surface with heterogeneous reactivity.  The
propagator of a diffusion-reaction process with eventual absorption on
the heterogeneous partially reactive surface is expressed in terms of
a much simpler propagator toward a homogeneous perfectly reactive
surface.  In other words, the original problem with general Robin
boundary condition that includes in particular mixed Robin-Neumann
condition, is reduced to that with Dirichlet boundary condition.
Chemical kinetics on the surface is incorporated as a matrix
representation of the surface reactivity in the eigenbasis of the
Dirichlet-to-Neumann operator.  New spectral representations of
important characteristics of diffusion-controlled reactions, such as
the survival probability, the distribution of reaction times, and the
reaction rate, are deduced.  Theoretical and numerical advantages of
this spectral approach are illustrated by solving interior and
exterior problems for a spherical surface that may describe either an
escape from a ball or hitting its surface from outside.  The effect of
continuously varying or piecewise constant surface reactivity
(describing, e.g., many reactive patches) is analyzed.
\end{abstract}



\maketitle

\section{Introduction}
\label{sec:intro}

Marian von Smoluchowski first emphasized the importance of diffusive
dynamics of reactant molecules and thus laid the foundations for the
modern theory of diffusion-controlled reactions \cite{Smoluchowski17}.
In the basic description, the concentration $c(\x,t)$ of molecules,
diffusing toward a static catalytic surface with the diffusivity $D$,
obeys the diffusion equation in a bulk domain $\Omega$
\begin{equation}
\frac{\partial}{\partial t} c(\x,t) = D \Delta c(\x,t)  \qquad (\x\in \Omega)
\end{equation}
(with $\Delta$ being the Laplace operator), subject to the Dirichlet
boundary condition on the surface $\pa$:
\begin{equation} \label{eq:Dirichlet}
c(\x,t) = 0 \qquad (\x \in \pa).
\end{equation}
This condition describes a perfect sink, i.e., any molecule hitting
the surface reacts with infinite reaction rate upon the first
encounter.  Since that seminal paper by Smoluchowski,
diffusion-controlled reactions to perfect sinks and the related
first-passage phenomena have been thoroughly investigated
\cite{Rice85,Redner,Levitz06,Condamin07,Benichou10b,Benichou14,Metzler}.

The assumption of infinite reaction rate is not realistic for most
chemical reactions because a molecule that approached a catalytic
surface, needs to overcome an activation energy barrier to react that
results in a finite reaction rate \cite{Weiss86,Hanggi90}.  This
effect was first incorporated by Collins and Kimball \cite{Collins49}
who replaced Dirichlet boundary condition (\ref{eq:Dirichlet}) by
Robin boundary condition (also known as Fourier, radiation, or third
boundary condition):
\begin{equation} \label{eq:Robin}
 - D \frac{\partial}{\partial \n} c(\x,t) = \kappa(\x) \, c(\x,t)  \qquad (\x \in \pa),
\end{equation}
where $\partial/\partial \n$ is the normal derivative oriented outward
the bulk.  This condition states that at each boundary point $\s \in
\pa$, the diffusive flux density, $j = (-D \nabla c) \cdot \n_{\s}$,
in the unit direction $\n_{\s}$ orthogonal to the surface, is
proportional to the concentration at this point.  The proportionality
coefficient, $\kappa(\s)$, is called the reactivity (with units m/s)
and can in general depend on the point $\s$.  In this formulation,
Robin boundary condition is essentially a mass conservation law at
each point of the boundary: the net influx of molecules diffusing
toward the boundary is equal to the amount of reacted molecules.  The
limit $\kappa(\s) = 0$ (for all $\s$) describes an inert surface
without any reaction (i.e., the net diffusive flux at the surface is
zero), whereas the limit $\kappa(\s) = \kappa \to \infty$ reduces
Eq. (\ref{eq:Robin}) to Eq. (\ref{eq:Dirichlet}) and describes an
immediate reaction upon the first encounter.  The reactivity is thus
related to the probability of reaction event at the encounter
\cite{Filoche99,Grebenkov03,Grebenkov06a}.  Robin boundary
condition with homogeneous (constant) reactivity $\kappa$ was often
employed to describe many chemical and biochemical reactions and
permeation processes
\cite{Lauffenburger,Sano79,Sano81,Shoup82,Sapoval94,Sapoval02,Grebenkov05,Qian06,Traytak07,Bressloff08,Galanti16,Grebenkov19b},
to model stochastic gating
\cite{Benichou00,Reingruber09,Lawley15,Bressloff17}, or to approximate
the effect of microscopic heterogeneities in a random distribution of
reactive sites \cite{Berg77,Shoup81,Zwanzig91} (see a recent overview
in \cite{Grebenkov19}).

In spite of its practical importance, diffusion-controlled reactions
with heterogeneous surface reactivity $\kappa(\s)$ remain much less
studied.  In fact, when $\kappa(\s)$ is not constant, the
eigenfunctions of the Laplace operator with Robin boundary condition
are not known explicitly even for simple domains (e.g. a ball) that
prohibits using standard spectral decompositions \cite{Carslaw,Crank},
on which most of classical solutions are based.  One needs therefore
to resort to numerical tools such as finite element or finite
difference methods for solving the diffusion equation, or to Monte
Carlo simulations.  A notable exception is the case of piecewise
constant reactivity that describes a target $\Gamma$ (or multiple
targets $\Gamma_i$) with a constant reactivity $\kappa$ on the
otherwise inert surface.  This situation corresponds to Robin-Neumann
(for $0 < \kappa < \infty$) or Dirichlet-Neumann (for $\kappa =
\infty$) mixed boundary conditions \cite{Sneddon,Duffy}.  The
Dirichlet-Neumann boundary value problem has been particularly well
studied for the Poisson and Laplace equations determining the mean
first-passage time to a small target and the reaction rate,
respectively (see overviews in
\cite{Holcman14,Schuss,Holcman} and references therein).  On one hand,
matched asymptotic analysis, dual series technique, and conformal
mapping were applied to establish the behavior of the mean
first-passage time in both two- and three-dimensional domains
\cite{Singer06a,Singer06b,Singer06c,Pillay10,Cheviakov10,Cheviakov12,Caginalp12,Marshall16,Grebenkov16c}.
On the other hand, homogenization techniques were used to substitute
piecewise constant reactivity $\kappa(\s)$ by an effective homogeneous
reactivity 
\cite{Berg77,Shoup81,Zwanzig90,Zwanzig91,Berezhkovskii04,Berezhkovskii06,Muratov08,Bernoff18b,Dagdug16,Lindsay17,Bernoff18a}.
More recent works investigated how the mean reaction time is affected
by a finite lifetime of diffusing particles
\cite{Yuste13,Meerson15,Grebenkov17d}, by partial reactivity and
interactions \cite{Grebenkov17a,Agranov18}, by target aspect ratio
\cite{Grebenkov17b}, by reversible target-binding kinetics
\cite{Grebenkov17c,Lawley19} and surface-mediated diffusion
\cite{Benichou10,Benichou11,Rupprecht12a,Rupprecht12b}, by
heterogeneous diffusivity \cite{Vaccario15}, and by rapid
re-arrangments of the medium \cite{Jain16,Lanoiselee18,Sposini19}.
Some of the related effects onto the whole distribution of reaction
times were analyzed 
\cite{Godec16a,Godec16b,Grebenkov18,Grebenkov18c,Hartich18}.  However,
the current understanding of diffusion-controlled reactions on
catalytic surfaces with continuously varying heterogeneous reactivity
remains episodic.

In this paper, we propose a mathematical description of
diffusion-controlled reactions on catalytic surfaces, in which
chemical kinetics, characterized by heterogeneous surface reactivity
$\kappa(\s)$, is disentangled from the first-passage diffusive steps.
In Sec. \ref{sec:theory}, we express the propagator of the
sophisticated diffusion-reaction process with multiple reflections on
partially reactive surface in terms of a much simpler Dirichlet
propagator toward a homogeneous perfectly reactive surface with
Dirichlet boundary condition.  Chemical kinetics is incorporated via a
matrix representation of the heterogeneous surface reactivity in the
eigenbasis of the Dirichlet-to-Neumann operator, which is also tightly
related to the Dirichlet propagator.  From the propagator, we deduce
other important characteristics of diffusion-controlled reactions such
as the survival probability, the distribution of reaction times and
the reaction rate.  This formalism provides a general description of
such processes and brings conceptually new tools for its
investigation.  In Sec. \ref{sec:sphere}, this spectral approach is
applied to an important example of a spherical surface for which the
Dirichlet propagator and the Dirichlet-to-Neumann operator are known
explicitly.  We study both the interior and exterior problems that may
describe either an escape from a ball or hitting its surface from
outside.  Semi-analytical solutions for the probability density of
reaction times and for the reaction rate are derived.  In
Sec. \ref{sec:discussion}, we discuss the advantages and limitations
of the spectral approach, its possible extensions, and further
applications, in particular, for analytical and numerical studies of
mixed boundary value problems.  Technical derivations are reported in
Appendices.

\section{General spectral description}
\label{sec:theory}

We consider a molecule diffusing with the diffusion coefficient $D$ in
an Euclidean domain $\Omega \subset\R^d$ toward a partially reactive
catalytic boundary $\pa$ characterized by a prescribed heterogeneous
(space-dependent) reactivity $0 \leq \kappa(\s) < \infty$.  Once the
molecule hits the boundary at some point $\s$, it may either react or
be reflected back to resume its diffusion until the next encounter,
and so on.  The reaction probability at each encounter is
characterized by the reactivity $\kappa(\s)$ at the encounter point.
In this way, the molecule performs multiple diffusive excursions in
the bulk until reaction occurs.  The finite reactivity results
therefore in a very sophisticated diffusive dynamics near the
catalytic surface, which is much more intricate than just the first
arrival to a homogeneous perfectly reactive surface.  A probabilistic
construction of this diffusive process (called partially reflected
Brownian motion) was discussed in
\cite{Grebenkov03,Grebenkov06a,Papanicolaou90,Bass08,Grebenkov07a,Singer08,Grebenkov09}
(see an overview in \cite{Grebenkov19}).

Without dwelling on the probabilistic aspects of the problem, we aim
at characterizing such diffusion-reaction processes via the {\it
propagator} $G(\x,t|\x_0)$ (also known as heat kernel or Green's
function).  This is the probability density for a molecule that has
not reacted until time $t$ on the partially reactive boundary $\pa$,
to be in a vicinity of a point $\x$ at time $t$, given that it was
started at a point $\x_0$ at time $0$.  For any fixed starting point
$\x_0 \in \overline{\Omega} = \Omega \cup \pa$, the propagator
satisfies the following boundary value problem
\begin{subequations}  \label{eq:Gtime}
\begin{eqnarray}  
\frac{\partial G(\x,t|\x_0)}{\partial t} - D \Delta  G(\x,t|\x_0) &=& 0  \quad (\x \in \Omega), \\
G(\x,t=0|\x_0) &=& \delta(\x - \x_0) , \\ \label{eq:G_Robin}
\biggl(D \frac{\partial}{\partial \n_{\x}} + \kappa(\x)\biggr)  G(\x,t|\x_0) &=& 0  \quad (\x \in \pa),  
\end{eqnarray}
\end{subequations}
where $\delta(\x-\x_0)$ is the Dirac distribution, and the Laplace
operator $\Delta$ acts on $\x$.  If the domain $\Omega$ is unbounded,
these equations are completed by the regularity condition at infinity:
$G(\x,t|\x_0) \to 0$ as $|\x|\to\infty$.  To avoid technicalities, we
assume that the boundary $\pa$ is smooth.  The following discussion
extends our former results
\cite{Grebenkov06a,Grebenkov06,Grebenkov07a,Grebenkov19} to
heterogeneous reactivity and time-dependent diffusion equation.

We consider the Laplace-transformed propagator,
\begin{equation}
\tilde{G}(\x,p|\x_0) = \int\limits_0^\infty dt \, e^{-pt} \, G(\x,t|\x_0),
\end{equation}
which satisfies the modified Helmholtz equation for each fixed $\x_0
\in \overline{\Omega}$:
\begin{subequations}  \label{eq:tildeG_eq0}
\begin{eqnarray}  \label{eq:tildeG_eq}
(p - D \Delta) \tilde{G}(\x,p|\x_0) &=& \delta(\x - \x_0), \\  
\biggl(D \frac{\partial }{\partial \n_{\x}} + \kappa(\x)\biggr) \tilde{G}(\x,p|\x_0)  &=& 0 \quad (\x \in \pa) 
\end{eqnarray}
\end{subequations}
(tilde will denote Laplace-transformed quantities).

Our goal is to express the propagator $\tilde{G}(\x,p|\x_0)$
describing diffusion toward heterogeneous partially reactive surface
$\pa$ in terms of the much simpler Dirichlet propagator
$\tilde{G}_0(\x,p|\x_0)$ that characterizes diffusion toward the {\it
homogeneous perfectly} reactive surface and satisfies for each fixed
$\x_0 \in \overline{\Omega}$:
\begin{subequations}  \label{eq:tildeG0}
\begin{eqnarray} 
(p - D \Delta) \tilde{G}_0(\x,p|\x_0) &=& \delta(\x - \x_0), \\ 
\tilde{G}_0(\x,p|\x_0)  &=& 0 \quad (\x \in \pa) .
\end{eqnarray}
\end{subequations}

Due to the linearity of the problem (\ref{eq:tildeG_eq0}), one can
search its solution in the form
\begin{equation}   \label{eq:g_reg}
\tilde{G}(\x,p|\x_0) = \tilde{G}_0(\x,p|\x_0) + \tilde{g}(\x,p|\x_0), 
\end{equation}
where the unknown regular part $\tilde{g}(\x,p|\x_0)$ satisfies
\begin{subequations} \label{eq:gtilde_eq}
\begin{eqnarray}  
(p - D \Delta) \tilde{g}(\x,p|\x_0) &=& 0  \quad (\x\in\Omega), \\ 
\biggl(D \frac{\partial}{\partial \n_{\x}} + \kappa(\x) \biggr) \tilde{g}(\x,p|\x_0) &=& \tilde{j}_0(\x,p|\x_0) 
\end{eqnarray}
\end{subequations}
for $\x \in \pa$, where
\begin{equation}
\tilde{j}_0(\s,p|\x_0) = - D \left. \biggl( \frac{\partial}{\partial \n_{\x}} \tilde{G}_0(\x,p|\x_0) \biggr) \right|_{\x = \s}  \quad (\s\in\pa)
\end{equation}
is the Laplace transform of the diffusive flux density
$j_0(\s,t|\x_0)$ at time $t$ in a point $\s$ of the homogeneous
perfectly reactive surface (i.e., the probability density of the first
arrival in a vicinity of $\s$ at time $t$ after starting from $\x_0$
at time $0$).

\subsection{Dirichlet-to-Neumann operator}

The solution of the boundary value problem (\ref{eq:gtilde_eq}) can be
obtained with the help of the {\it Dirichlet-to-Neumann operator}
$\M_p$ (also known as Poincar\'e-Steklov operator)
\cite{Egorov,Jacob,Taylor}.  This is a pseudo-differential
self-adjoint operator that associates to a function $\tilde{f}$ on the
boundary $\pa$ another function on that boundary:
\begin{equation}
[\M_p \tilde{f}](\s) = \left. \biggl(\frac{\partial \tilde{u}(\x,p)}{\partial \n}\biggr) \right|_{\x = \s}  \quad (\s\in\pa),
\end{equation}
where $\tilde{u}(\x,p)$ is the solution of the Dirichlet boundary
value problem:
\begin{subequations}  \label{eq:u_Dirichlet}
\begin{eqnarray}
(p - D \Delta) \tilde{u}(\x,p) &=& 0 \hskip 12mm (\x\in\Omega), \\
\tilde{u}(\x,p) &=& \tilde{f}(\x,p) \quad (\x\in\pa) 
\end{eqnarray}
\end{subequations}
(here we skip the usual regularity assumptions on $\pa$, as well as
the explicit description of the functional spaces involved in the
rigorous definition of $\M_p$, see
\cite{Egorov,Jacob,Taylor,Marletta04,Arendt07,Arendt15,Hassell17} for
details).  For instance, if $\tilde{f}$ is understood as a source of
molecules on the boundary $\pa$ emitted into the reactive bulk, then
the operator $\M_p$ gives their flux density on that boundary.  Note
that there is a family of operators parameterized by $p$ (or $p/D$).

As the solution of the Dirichlet boundary value problem
(\ref{eq:u_Dirichlet}) can be expressed in terms of the Dirichlet
propagator $\tilde{G}_0(\x,p|\x_0)$ in a standard way,
\begin{equation*}
\tilde{u}(\x,p) = \int\limits_\pa d\s' \,  \tilde{j}_0(\s',p|\x) \, \tilde{f}(\s',p),
\end{equation*}
the Dirichlet-to-Neumann propagator acts formally as
\begin{equation}  \label{eq:Mp_j0}
[\M_p \tilde{f}](\s) = \left. \left(\frac{\partial}{\partial \n} \int\limits_{\pa} d\s' \,
\tilde{j}_0(\s',p|\x)\, \tilde{f}(\s',p) \right)\right|_{\x = \s} ,
\end{equation}
and thus the Dirichlet propagator determines the Dirichlet-to-Neumann
operator $\M_p$.  In Appendix \ref{sec:ADirichlet}, it is also shown
how the Dirichlet propagator can be constructed from the operator
$\M_p$.  As a consequence, these two important objects are equivalent.
As discussed in \cite{Grebenkov06a} for the case $p = 0$, the
Dirichlet-to-Neumann operator can also be interpreted as the
continuous limit of the Brownian self-transport operator $Q_{ij}$
which was introduced in \cite{Filoche99,Grebenkov03} to describe the
probability of the first arrival to a site $j$ of a discretized
boundary from another site $i$ via bulk diffusion.

Let us now return to the boundary value problem (\ref{eq:gtilde_eq}).
Suppose that we have solved this problem and found that the solution
$\tilde{g}(\x,p|\x_0)$ on the boundary $\pa$ is equal to some function
$\tilde{f}(\x,p)$.  Applying then the Dirichlet-to-Neumann operator to
$\tilde{f}(\x,p)$, one can express the normal derivative of
$\tilde{g}(\x,p|\x_0)$, from which
\begin{equation}
\tilde{g}(\s,p|\x_0) = \bigl(\M_p + \K \bigr)^{-1} \, \frac{\tilde{j}_0(\s,p|\x_0)}{D}  \quad (\s\in\pa),
\end{equation}
where $\K$ is the operator of multiplication by $\kappa(\s)/D$.
Knowing the restriction of $\tilde{g}(\x,p|\x_0)$ on the boundary
$\pa$, one can reconstruct this function in the bulk $\Omega$ as the
solution of the corresponding Dirichlet problem:
\begin{equation}  
\tilde{g}(\x,p|\x_0) = \int\limits_{\pa} d\s \, \tilde{j}_0(\s,p|\x)\, \tilde{g}(\s,p|\x_0)  .
\end{equation}
In this way, we obtain the desired representation of the propagator in
the form of a scalar product between two functions on the boundary
\begin{eqnarray}  \label{eq:Gtilde}
&& \tilde{G}(\x,p|\x_0) = \tilde{G}_0(\x,p|\x_0) \\  \nonumber
&& + \frac{1}{D} \biggl( \tilde{j}_0(\cdot,p|\x)  
\, \cdot \, (\M_p + \K)^{-1} \tilde{j}_0(\cdot,p|\x_0) \biggr)_{L_2(\pa)} ,
\end{eqnarray}
where $(f \cdot g)_{L_2(\pa)}$ denotes the standard scalar
product between functions $f$ and $g$ on the boundary $\pa$:
\begin{equation*}
(f \cdot g)_{L_2(\pa)} = \int\limits_\pa d\s \, f(\s) \, g^*(\s),
\end{equation*}
and asterisk denotes the complex conjugate.  Equation
(\ref{eq:Gtilde}) is the first main result of the paper.  Remarkably,
all the ``ingredients'' of this formula correspond to the Dirichlet
condition on a homogeneous perfectly reactive boundary, except for the
operator $\K$ that keeps track of heterogeneous surface reactivity
$\kappa(\s)$.  We outline that Eq. (\ref{eq:Gtilde}) does not solve
the original problem but reduces it to a much simpler and more
thoroughly studied Dirichlet problem.

When $\x$ and $\x_0$ are boundary points, the identity
$\tilde{j}_0(\s,p|\s_0) = \delta(\s - \s_0)$ reduces
Eq. (\ref{eq:Gtilde}) to
\begin{equation}  \label{eq:resolvent}
D \tilde{G}(\s,p|\s_0) = \bigl(\M_p + \K\bigr)^{-1}  \delta(\s - \s_0) \quad (\s_0,\s\in\pa),
\end{equation}
i.e., $D \tilde{G}(\s,p|\s_0)$ is the kernel of the operator $\M_p +
\K$.  One can therefore rewrite Eq. (\ref{eq:Gtilde}) as
\begin{eqnarray}   \label{eq:Gprob_Laplace}
&& \tilde{G}(\x,p|\x_0) = \tilde{G}_0(\x,p|\x_0) \\    \nonumber
&+& \int\limits_\pa d\s_1 \int\limits_\pa d\s_2  \, \tilde{j}_0(\s_1,p|\x_0)  \,
\tilde{G}(\s_2,p|\s_1) \, \tilde{j}_0(\s_2,p|\x) ,
\end{eqnarray}
while its inverse Laplace transform reads
\begin{eqnarray}    \nonumber
&& G(\x,t|\x_0) = G_0(\x,t|\x_0) + \int\limits_\pa d\s_1 \int\limits_\pa d\s_2 \int\limits_0^t dt_1 \int\limits_{t_1}^t dt_2 \\    \label{eq:Gprob} 
&& \times  j_0(\s_1,t_1|\x_0)  \, G(\s_2,t_2-t_1|\s_1) \, j_0(\s_2,t-t_2|\x) .
\end{eqnarray}
This relation expresses the propagator $G(\x,t|\x_0)$ in the whole
domain in terms of the propagator $G(\s_2,t|\s_1)$ from one boundary
point to another boundary point via bulk diffusion.  The first term
represents the contribution of direct trajectories from $\x_0$ to $\x$
that do not touch the boundary $\pa$.  The second term also has a
simple probabilistic interpretation: a molecule reaches the boundary
for the first time at $t_1$, performs partially reflected Brownian
motion over time $t_2-t_1$ (with eventual failed attempts of reaction
at each encounter with the surface), and diffuses to the bulk point
$\x$ during time $t-t_2$ without hitting the reactive surface.

When $\x = \s$ is a boundary point, one has $G_0(\s,t|\x_0) = 0$ and
$j_0(\s_2, t-t_2|\s) = \delta(\s-\s_2) \delta(t-t_2)$, so that the
integrals over $\s_2$ and $t_2$ are removed, reducing
Eq. (\ref{eq:Gprob}) to
\begin{equation}  \label{eq:Gprob_s}
G(\s,t|\x_0) = \int\limits_\pa d\s_1 \int\limits_0^t dt_1 \, j_0(\s_1,t_1|\x_0) \, G(\s,t-t_1|\s_1).
\end{equation}
This relation justifies the qualitative separation of the
diffusion-reaction process into two steps: the first arrival step
(described by $j_0(\s_1,t_1|\x_0)$) and the reaction step (described
by $G(\s,t-t_1|\s_1)$).  We stress, however, that the reaction step
involves intricate diffusion process near the partially reactive
catalytic surface.  In addition to the new conceptual view onto
partially reflected Brownian motion, the representations
(\ref{eq:Gprob}, \ref{eq:Gprob_s}) can be helpful for a numerical
computation of the propagator because only the boundary-to-boundary
transport via $G(\s_2,t|\s_1)$ needs to be determined.  This kernel
significantly extends the Brownian self-transport operator introduced
in \cite{Filoche99,Grebenkov03} (see below).

\subsection{Other common diffusion characteristics}

The propagator $G(\x,t|\x_0)$ determines many quantities often
considered in the context of diffusion-controlled reactions such as
the survival probability up to time $t$, the reaction time
distribution, the distribution of reaction points (at which reaction
occurs), and the reaction rate.  For instance, the diffusive flux
density at a partially reactive point $\s \in \pa$ is
\begin{equation}  \label{eq:jt}
j(\s,t|\x_0) = \left. \biggl( - D\frac{\partial G(\x,t|\x_0)}{\partial \n_{\x}} \biggr) \right|_{\x=\s} \hspace*{-1mm} = \kappa(\s) G(\s,t|\x_0) ,
\end{equation}
where we used the Robin boundary condition (\ref{eq:G_Robin}).  This
is the joint probability density for the reaction time and the
reaction point on the catalytic surface.  The integral over $\s$
yields the marginal probability density of reaction times,
\begin{equation}  \label{eq:rhot}
H(t|\x_0) = \int\limits_\pa d\s \, j(\s,t|\x_0) = \int\limits_\pa d\s \, \kappa(\s) \, G(\s,t|\x_0) ,
\end{equation}
whereas the integral over $t$ gives the marginal probability density
of reaction points:
\begin{equation}  \label{eq:omega}
\omega(\s|\x_0) = \int\limits_0^\infty dt \, j(\s,t|\x_0) = \tilde{j}(\s,0|\x_0) = \kappa(\s) \tilde{G}(\s,0|\x_0).
\end{equation}
The latter was called the {\it spread harmonic measure} density
\cite{Grebenkov06a,Grebenkov06,Grebenkov06b,Grebenkov15}.  This is a
natural extension of the harmonic measure density
$\tilde{j}_0(\s,0|\x_0)$ that characterizes the first arrival onto the
perfectly reactive surface \cite{Garnett,Grebenkov05a,Grebenkov05b}.
As the probability density $H(t|\x_0)$ can be interpreted as the
probability flux onto the surface for a molecule started from $\x_0$,
its integral with the initial concentration of molecules, $c_0(\x_0)$,
yields the overall diffusive flux onto the surface, i.e., the reaction
rate:
\begin{equation}  \label{eq:kt}
J(t) = \int\limits_\Omega d\x_0 \, c_0(\x_0) \, H(t|\x_0).
\end{equation}
In turn, the integral of $H(t|\x_0)$ from $t$ to infinity gives the
survival probability up to time $t$:
\begin{equation}
S(t|\x_0) = 1 - \int\limits_0^t dt' \, H(t'|\x_0) ,
\end{equation}
while $1 - S(t|\x_0)$ is the probability of reaction up to time $t$.
All these quantities are expressed in terms of the propagator and thus
determined from Eq. (\ref{eq:Gtilde}).

\subsection{Spectral decompositions}
\label{sec:spectral}

When the boundary $\pa$ is bounded, the Dirichlet-to-Neumann operator
$\M_p$ has a discrete spectrum, with a set of nonnegative eigenvalues
$\mu_n^{(p)}$ and $L_2(\pa)$-normalized eigenfunctions $v_n^{(p)}$
forming a complete orthogonal basis in $L_2(\pa)$:
\begin{equation}
\M_p v_n^{(p)}(\s) = \mu_n^{(p)} v_n^{(p)}(\s)  \qquad  (n = 0,1,\ldots).
\end{equation}
We emphasize that both $\mu_n^{(p)}$ and $v_n^{(p)}$ depend in general
on $p$ as a parameter.  Expanding the scalar product in
Eq. (\ref{eq:Gtilde}) over this basis, one gets
\begin{eqnarray}  \label{eq:Gtilde2}
&& \tilde{G}(\x,p|\x_0) = \tilde{G}_0(\x,p|\x_0) \\  \nonumber
&& + \frac{1}{D} \sum\limits_{n,n'=0}^\infty V_n^{(p)}(\x_0) 
\bigl[(\MM + \KK)^{-1}\bigr]_{n,n'} [V_{n'}^{(p)}(\x)]^* ,
\end{eqnarray}
where
\begin{equation}  \label{eq:Vnp}
V_n^{(p)}(\x_0) = \int\limits_\pa d\s \, \tilde{j}_0(\s,p|\x_0) \, v_n^{(p)}(\s)
\end{equation}
is the projection of the Laplace-transformed flux density
$\tilde{j}_0(\s,p|\x_0)$ onto the eigenfunction $v_n^{(p)}(\s)$, and
\begin{subequations}  
\begin{eqnarray} \label{eq:MK_def}
\MM_{n,n'} &=& \delta_{nn'} \mu_n^{(p)} , \\  \label{eq:MK_def2}
\KK_{n,n'} &=& \int\limits_\pa d\s \, [v_n^{(p)}(\s)]^* \, \frac{\kappa(\s)}{D} \, v_{n'}^{(p)}(\s) 
\end{eqnarray}
\end{subequations}
are infinite-dimensional matrices that represent the
Dirichlet-to-Neumann operator $\M_p$ and the reactivity multiplication
operator $\K$ in the basis of eigenfunctions $v_n^{(p)}(\s)$.

From the spectral representation (\ref{eq:Gtilde2}) and
Eq. (\ref{eq:jt}), we deduce
\begin{equation}  \label{eq:jtilde}
\tilde{j}(\s,p|\x_0) = \sum\limits_{n,n'=0}^\infty V_n^{(p)}(\x_0) 
\bigl[(\MM + \KK)^{-1} \KK \bigr]_{n,n'} [v_{n'}^{(p)}(\s)]^* ,
\end{equation}
where we used the completeness of eigenfunctions $v_n^{(p)}$ to
represent $\kappa(\s)/D$ as multiplication by the matrix $\KK$.
According to Eqs. (\ref{eq:rhot}, \ref{eq:omega}), the spectral
decomposition (\ref{eq:jtilde}) yields immediately
\begin{equation}  \label{eq:omega_spectral}
\omega(\s|\x_0) = \sum\limits_{n,n'=0}^\infty V_n^{(0)}(\x_0) 
\bigl[(\MM + \KK)^{-1} \KK\bigr]^{(p=0)}_{n,n'}  \, [v_{n'}^{(0)}(\s)]^* 
\end{equation}
and
\begin{equation}  \label{eq:Htilde}
\tilde{H}(p|\x_0) = |\pa|^{1/2} \sum\limits_{n=0}^\infty h_n^{(p)} \, V_n^{(p)}(\x_0)  ,
\end{equation}
where
\begin{equation}  \label{eq:hn}
h_n^{(p)} = |\pa|^{-1/2} \sum\limits_{n'=0}^\infty \bigl[(\MM + \KK)^{-1} \KK \bigr]_{n,n'} \int\limits_\pa d\s \, [v_{n'}^{(p)}(\s)]^* 
\end{equation}
are dimensionless coefficients.  In particular, $\tilde{H}(0|\x_0)$ is
the reaction probability (in Appendix \ref{sec:reacprob}, we prove the
expected identity $\tilde{H}(0|\x_0) = 1$ for any bounded domain).
According to Eq. (\ref{eq:kt}), the Laplace-transformed reaction rate
is then
\begin{equation}  \label{eq:Ktilde}
\tilde{J}(p) = |\pa|^{1/2} \sum\limits_{n,n'=0}^\infty h_n^{(p)} \int\limits_\Omega d\x_0 \, V_n^{(p)}(\x_0) \, c_0(\x_0) .
\end{equation}
In Appendix \ref{sec:AVn}, we show how this expression can be further
simplified in the case of the uniform initial concentration.

While we mainly focus on Laplace-transformed quantities, their
representations in time domain can be obtained via Laplace transform
inversion either analytically or numerically.  For instance, the
inversion in the case of bounded domains can be performed via the
residue theorem by computing the poles $\{p_n\} \subset \C$ of
functions in Eqs. (\ref{eq:Gtilde2}, \ref{eq:Htilde},
\ref{eq:Ktilde}), which are determined by the condition
\begin{equation} \label{eq:det}
\det(\MM + \KK) = 0.
\end{equation}

In general, the spectral representation (\ref{eq:Gtilde2}) is not
simpler than Eq. (\ref{eq:Gtilde}) because all $V_n^{(p)}$, $\MM$ and
$\KK$ depend on $p$ as a parameter.  However, in some domains, these
``ingredients'' can be evaluated explicitly, providing a
semi-analytical form of the Laplace-transformed propagator and related
quantities.  We will illustrate this point in Sec. \ref{sec:sphere}
for a spherical boundary.

\subsection{Homogeneous partial reactivity}

In the particular case of homogeneous reactivity, $\kappa(\s) =
\kappa$, the operator $\K$ is proportional to the identity operator, and
Eq. (\ref{eq:resolvent}) implies that $D \tilde{G}(\s,p|\s_0)$ is the
resolvent of the Dirichlet-to-Neumann operator $\M_p$.  Moreover, as
\begin{equation} \label{eq:KK_hom}
\KK_{n,n'} = \delta_{n,n'}\, \frac{\kappa}{D} \,, 
\end{equation}
Eq. (\ref{eq:Gtilde2}) is reduced to
\begin{equation}  \label{eq:Gtilde2_hom}
\tilde{G}_{\rm hom}(\x,p|\x_0) = \tilde{G}_0(\x,p|\x_0) + \sum\limits_{n=0}^\infty 
\frac{V_n^{(p)}(\x_0) \, [V_n^{(p)}(\x)]^*}{D \mu_n^{(p)} + \kappa} \,.
\end{equation}
In turn, the condition (\ref{eq:det}) on the poles is reduced to a set
of decoupled equations

\begin{equation}  \label{eq:det_hom}
\mu_n^{(p)} + \frac{\kappa}{D} = 0 ,
\end{equation}
showing how the eigenvalues $\mu_n^{(p)}$ of the Dirichlet-to-Neumann
operator determine the eigenvalues of the associated Laplace operator
with Robin boundary condition.  

The other spectral decompositions are also simplified:
\begin{equation}  \label{eq:omega2_hom}
\omega_{\rm hom}(\s|\x_0) = \sum\limits_{n=0}^\infty \frac{V_n^{(0)}(\x_0) \, 
[v_n^{(0)}(\s)]^*}{\frac{D}{\kappa} \mu_n^{(0)} + 1} \,,
\end{equation}
\begin{equation}  \label{eq:Htilde2_hom}
\tilde{H}_{\rm hom}(p|\x_0) = \sum\limits_{n=0}^\infty \frac{V_n^{(p)}(\x_0) \, 
\int\nolimits_\pa d\s \, [v_n^{(p)}(\s)]^*}{\frac{D}{\kappa} \mu_n^{(p)} + 1} \,.
\end{equation}
and 
\begin{equation}  \label{eq:Jtilde2_hom}
\tilde{J}_{\rm hom}(p) = \frac{c_0 D}{p} \sum\limits_{n=0}^\infty 
\frac{\mu_n^{(p)} }{\frac{D}{\kappa} \mu_n^{(p)} + 1} \,  \left|\int\limits_\pa d\s \, v_n^{(p)}(\s) \right|^2  \,,
\end{equation}
where we used Eq. (\ref{eq:Ktilde2}) for the uniform initial
concentration $c_0$.  In the limit $p\to 0$, one recovers the formula
for the total steady-state flux derived in Ref. \cite{Grebenkov06};
Eq. (\ref{eq:Jtilde2_hom}) is therefore its extension to
time-dependent diffusion.  To our knowledge,
Eqs. (\ref{eq:Gtilde2_hom}, \ref{eq:omega2_hom}, \ref{eq:Htilde2_hom},
\ref{eq:Jtilde2_hom}) that are fully explicit in terms of the
eigenvalues and eigenfunctions of the Dirichlet-to-Neumann operator,
have not been earlier reported.  While alternative spectral
decompositions on the Laplace operator eigenfunctions are known for
bounded domains, there is no such expansion for unbounded domains, for
which the spectrum of the Laplace operator is continuous.  The
spectral formulation in terms of the eigenfunctions of the
Dirichlet-to-Neumann operator opens therefore new perspectives for
studying diffusion-reaction processes even for homogeneous reactivity.
From the numerical point of view, the computation of the
eigenfunctions of the Dirichlet-to-Neumann operator could in general
be simpler due to the reduced dimensionality: $v_n^{(p)}$ need to be
found on the boundary $\pa$, whereas the Laplace operator
eigenfunctions have to be computed in the whole domain $\Omega$.

\section{Spherical boundary}
\label{sec:sphere}

In this section, we apply our general spectral decompositions to the
case of a spherical boundary for which the eigenbasis of the
Dirichlet-to-Neumann operator is known explicitly.  We first discuss
in Sec. \ref{sec:interior} the interior problem that may describe, for
instance, an escape from a ball, and then in Sec. \ref{sec:exterior}
we dwell on the exterior problem and related chemical kinetics.  In
both cases, we provide semi-analytical solutions for an arbitrary
heterogeneous surface reactivity and then discuss some particular
cases, e.g., a piecewise constant reactivity that describes single or
multiple reactive targets on the otherwise inert boundary.  Technical
details of calculations are reported in Appendices
\ref{sec:Ainterior}, \ref{sec:K}, and \ref{sec:Aexterior}.

\subsection{Diffusion inside a ball}
\label{sec:interior}

We consider a diffusion-reaction process inside a ball of radius $R$,
$\Omega = \{ \x\in\R^3~:~ |\x| < R\}$, with a prescribed heterogeneous
surface reactivity $\kappa(\s)$.  For this domain, the eigenvalues and
eigenfunctions of the Dirichlet-to-Neumann operator are known
explicitly (Appendix \ref{sec:Ainterior}),
\begin{subequations}
\begin{eqnarray}
\mu_{nm}^{(p)} &=& \sqrt{p/D} \, \frac{i'_n(R\sqrt{p/D})}{i_n(R\sqrt{p/D})} \,, \\  \label{eq:vnm_sphere}
v_{nm}(\theta,\phi) &=& \frac{1}{R} \, Y_{mn}(\theta,\phi) ,
\end{eqnarray}
\end{subequations}
where $i_n(z)$ are the modified spherical Bessel functions of the
first kind, $Y_{mn}(\theta,\phi)$ are the $L_2(\pa)$-normalized
spherical harmonics, prime denotes the derivative with respect to the
argument, and we used spherical coordinates $(r,\theta,\phi)$.  Here
we employ the double index $nm$ to enumerate the eigenfunctions as
well as the elements of the matrices $\MM$ and $\KK$.  Note that the
eigenvalues do not depend on the index $m$ and thus are of
multiplicity $2n+1$, whereas the eigenfunctions $v_{nm}$ do not depend
on the parameter $p$.  The eigenvalues determine the matrix $\MM$ via
Eq. (\ref{eq:MK_def}), while Eq. (\ref{eq:MK_def2}) for the matrix
$\KK$ reads
\begin{equation}  \label{eq:KK_sphere}
\KK_{nm,n'm'} = \int\limits_0^\pi d\theta \, \sin\theta \int\limits_0^{2\pi} d\phi 
\frac{\kappa(\theta,\phi)}{D}  Y_{mn}^*(\theta,\phi)  Y_{m'n'}(\theta,\phi) .
\end{equation}

The calculation of the matrix $\KK$ in Eq. (\ref{eq:KK_sphere})
involves integrals with spherical harmonics that can often be
evaluated explicitly.  In Appendix \ref{sec:K}, we discuss several
common situations such as a single target, multiple non-overlapping
targets of circular shape or multiple latitudinal stripes,
axisymmetric reactivity $\kappa(\theta,\phi) = \kappa(\theta)$, and
an expansion of $\kappa(\theta,\phi)$ into a finite sum over spherical
harmonics.  Although cumbersome, resulting expressions for the matrix
$\KK$ are exact and do not involve numerical quadrature, providing a
powerful computational tool.  These cases can further be extended by
adding another concentric surface with reflecting or absorbing
boundary condition.  This modification does not change the matrix
$\KK$ but affects the eigenvalues of the Dirichlet-to-Neumann operator
and thus the matrix $\MM$.

As the Dirichlet propagator is also known, we deduce in Appendix
\ref{sec:Ainterior}
\begin{equation}
V_{nm}^{(p)}(\x_0) = R^{-1}  \, \frac{i_n(r_0\sqrt{p/D})}{i_n(R\sqrt{p/D})} \, Y_{mn}(\theta_0,\phi_0) \,,
\end{equation}
so that the Laplace-transformed propagator $\tilde{G}(\x,p|\x_0)$ is
determined in the semi-analytical form (\ref{eq:Gtilde2}), in which
the dependence on points $\x_0$ and $\x$ is fully explicit, whereas
the computation of the coefficients involves a numerical inversion of
the matrix $\MM + \KK$.  Similarly, one gets semi-analytical
expressions for the Laplace-transformed probability density of
reaction times and the spread harmonic measure (see Appendix
\ref{sec:Ainterior}), e.g.,
\begin{equation} \label{eq:Htilde_sphere_main}
\tilde{H}(p|\x_0) = \sqrt{4\pi} \sum\limits_{n=0}^\infty \sum\limits_{m=-n}^n h_{nm}^{(p)} \, 
\frac{i_n(r_0\sqrt{p/D})}{i_n(R\sqrt{p/D})} \, Y_{mn}(\theta_0,\phi_0),
\end{equation}
with
\begin{equation}  \label{eq:hnm}
h_{nm}^{(p)} = \bigl[(\MM + \KK)^{-1} \KK \bigr]_{nm,00} .
\end{equation}
The Laplace-transformed survival probability is related to
$\tilde{H}(p|\x_0)$ as 
\begin{equation}  \label{eq:Stilde}
\tilde{S}(p|\x_0) = \frac{1 - \tilde{H}(p|\x_0)}{p} \, ,
\end{equation}
whereas the mean reaction time is simply $\tilde{S}(0|\x_0)$.

\begin{figure}
\begin{center}
\includegraphics[width=27mm]{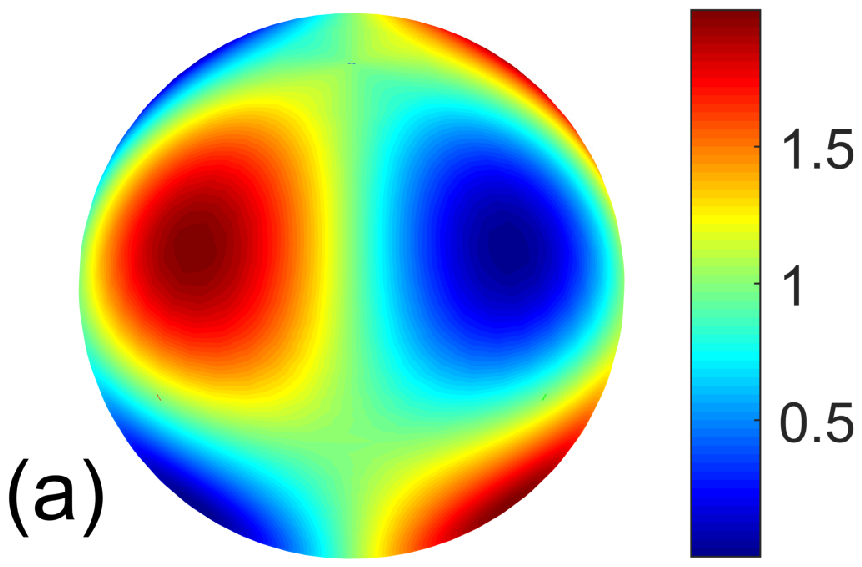} 
\includegraphics[width=27mm]{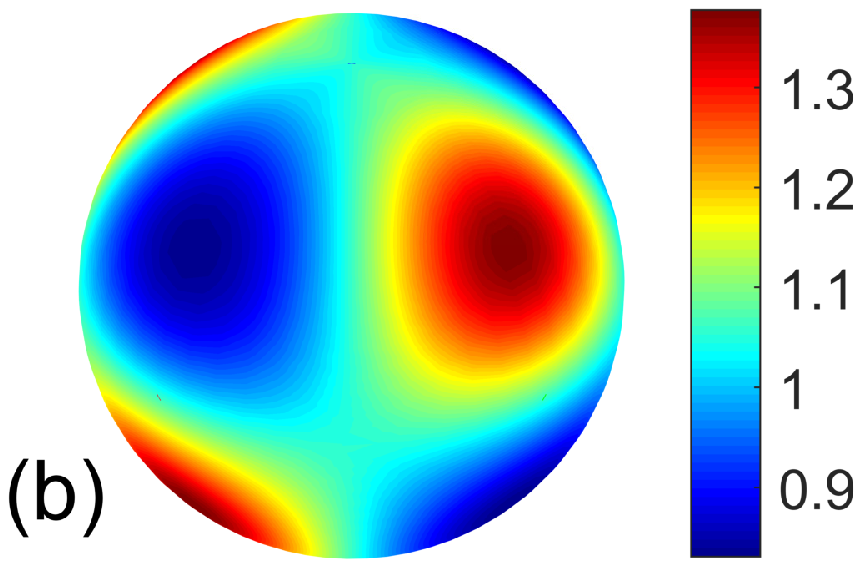} 
\includegraphics[width=27mm]{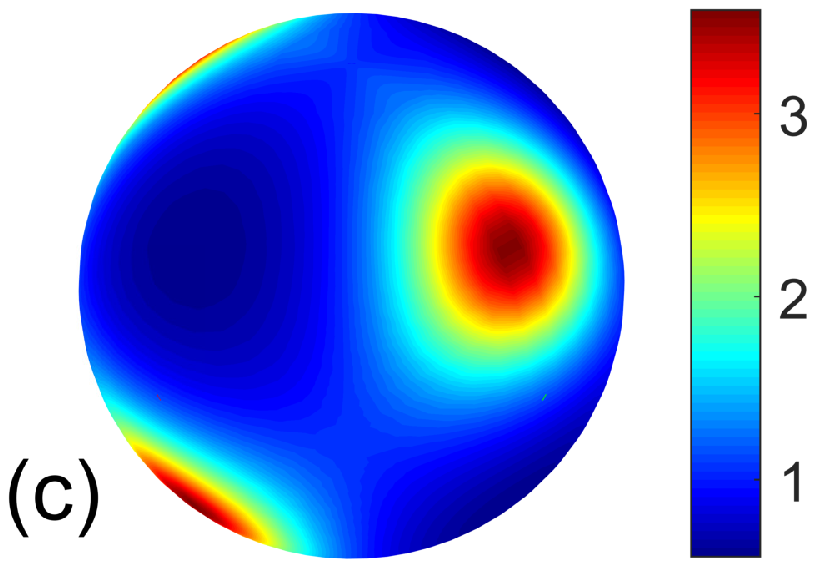} 
\end{center}
\caption{
Diffusion inside a ball of radius $R$.  {\bf (a)} Heterogeneous
surface reactivity $\kappa(\theta,\phi) = \kappa (1 + c
Y_{2,3}(\theta,\phi) + c Y_{-2,3}(\theta,\phi))$, with $\kappa R/D =
1$ and $c = 1.2728$ (this value ensures the positivity of
$\kappa(\theta,\phi)$).  {\bf (b,c)} Mean reaction time
$\tilde{S}(0|\x_0)$, rescaled by $\tilde{S}_{\rm hom}(0|\x_0) =
R/(3\kappa)$, as a function of the starting point $\x_0 =
(r_0,\theta_0,\phi_0)$ with $r_0 = R$, for such a reactivity
$\kappa(\theta,\phi)$, with $\kappa R/D = 1$ {\bf (b)} and $\kappa R/D
= 10$ {\bf (c)}.  The matrix $\KK$ was computed with the truncation
order $\nmax = 20$ as described in Appendix \ref{sec:K_general}.}
\label{fig:Tmean}
\end{figure}

Figures \ref{fig:Tmean}(b,c) illustrate how the mean reaction time
depends on the starting point $\x_0$ for a particular choice of a
continuously varying heterogeneous surface reactivity
$\kappa(\theta,\phi)$ shown in Fig. \ref{fig:Tmean}(a).  When the mean
reactivity is weak ($\kappa R/D = 1$, Fig. \ref{fig:Tmean}(b)),
$\tilde{S}(0|\x_0)$ is close to the mean reaction time $\tilde{S}_{\rm
hom}(0|\x_0) = R/(3\kappa)$ corresponding to homogeneous reactivity
$\kappa$.  Here, multiple failed reaction attempts homogenize the mean
reaction time, even though the starting point $\x_0$ lies on the
catalytic boundary.  In turn, significant deviations from
$R/(3\kappa)$ are observed at a larger mean reactivity $\kappa R/D =
10$.  In this case, the mean reactivity is not representative and
heterogeneities start to be more and more important.

\subsection{Diffusion outside a ball}
\label{sec:exterior}

For diffusion in the unbounded domain $\Omega = \{\x\in\R^3 ~:~ |\x| >
R\}$ outside the spherical surface of radius $R$, the eigenfunctions
of the Dirichlet-to-Neumann operator are still given by
Eq. (\ref{eq:vnm_sphere}), so that the matrix $\KK$ remains unchanged.
In turn, the matrix $\MM$ is now determined by the eigenvalues
\begin{equation}  \label{eq:mu_sphere_ext0}
\mu_{nm}^{(p)} = - \sqrt{p/D} \, \frac{k'_n(R\sqrt{p/D})}{k_n(R\sqrt{p/D})} \,, 
\end{equation}
where $k_n(z)$ are the modified spherical Bessel function of the
second kind.  From the known Dirichlet propagator, we compute in
Appendix \ref{sec:Aexterior}
\begin{equation}  \label{eq:Vnm_sphere_ext0}
V_{nm}^{(p)}(\x_0) = R^{-1}  \frac{k_n(r_0\sqrt{p/D})}{k_n(R\sqrt{p/D})} \, Y_{mn}(\theta_0,\phi_0) \,.
\end{equation}
As a consequence, our spectral decomposition (\ref{eq:Gtilde2}) fully
determines the Laplace-transformed propagator $\tilde{G}(\x,p|\x_0)$.
The Laplace-transformed probability density of reaction times is again
obtained from Eq. (\ref{eq:Htilde}):
\begin{equation} \label{eq:Htilde_sphere_ext}
\tilde{H}(p|\x_0) = \sqrt{4\pi}\sum\limits_{n=0}^\infty \sum\limits_{m=-n}^n h_{nm}^{(p)} \, 
\frac{k_n(r_0\sqrt{p/D})}{k_n(R\sqrt{p/D})} \, Y_{mn}(\theta_0,\phi_0),
\end{equation}
with $h_{nm}^{(p)}$ given by Eq. (\ref{eq:hnm}), in which the matrices
$\KK$ and $\MM$ are determined by Eqs. (\ref{eq:KK_sphere},
\ref{eq:mu_sphere_ext0}).  The Laplace-transformed survival
probability is still given by Eq. (\ref{eq:Stilde}), while the mean
reaction time is infinite.

According to Eq. (\ref{eq:Ktilde}), the Laplace-transformed reaction
rate can be obtained by integrating $\tilde{H}(p|\x_0)$ with the
initial concentration of molecules, which for the uniform
concentration, $c_0(\x_0) = c_0$, yields
\begin{equation} \label{eq:Ktilde_sphere}
\tilde{J}(p) = 4\pi D R c_0 \, \frac{R \mu_{00}^{(p)} \, h_{00}^{(p)}}{p} 
\end{equation}
(note that the same formula with the appropriate matrix $\MM$ holds
for diffusion inside the ball).  The prefactor $4\pi D R c_0$ is the
Smoluchowski rate to a homogeneous perfectly reactive ball of radius
$R$, whereas the second factor describes the effect of heterogeneous
surface reactivity.  In the limit $p\to 0$, this expression yields the
steady-state reaction rate
\begin{equation}  \label{eq:k_steady}
J(\infty) = 4\pi DR c_0 \, h_{00}^{(0)} .
\end{equation}
For the homogeneous reactivity, our formulas are reduced to that of
Collins and Kimball \cite{Collins49}, see Appendix
\ref{sec:Aexterior}, with
\begin{equation}  \label{eq:CK_infty}
h_{00}^{(0)} = \frac{1}{1 + D/(\kappa R)} \,.
\end{equation}

\begin{figure}
\begin{center}
\includegraphics[width=88mm]{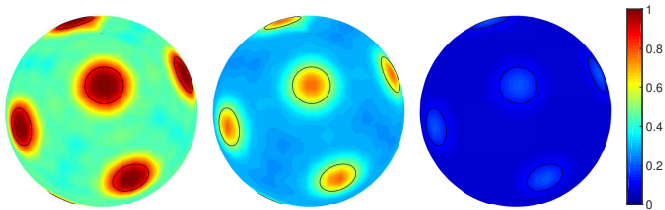}
\end{center}
\caption{
Diffusion outside a ball of radius $R$.  The reaction probability
$\tilde{H}(0|\x_0)$ on the spherical surface of radius $R$ as a
function of the starting point $\x_0 = (r_0,\theta_0,\phi_0)$ with
$r_0 = R$.  Ten circular targets of angular size $\ve = 0.2$ (shown by
thin black circles) are evenly distributed on the surface, with three
values of reactivity: $\kappa R/D = 100$ (left), $\kappa R/D = 10$
(middle), and $\kappa R/D = 1$ (right).  The matrix $\KK$ was computed
with the truncation order $\nmax = 20$ as described in Appendix
\ref{sec:multiple}. }
\label{fig:QNtarget}
\end{figure}

Figure \ref{fig:QNtarget} shows the reaction probability
$\tilde{H}(0|\x_0)$ from Eq. (\ref{eq:Htilde_sphere_ext}) (see also
Eq. (\ref{eq:H0_ext})) on the inert spherical surface covered by ten
evenly distributed circular partially reactive targets of angular size
$\ve = 0.2$.  Even though the starting point $\x_0$ lies on the
surface ($r_0 = R$), the reaction probability is not equal to $1$ due
to the partial reactivity and eventual failed attempts to react.  When
the reactivity is large ($\kappa R/D = 100$, left panel), the reaction
probability is close to $1$ when the molecule starts at any target,
and drops to $0.4$ in between two targets.  At intermediate reactivity
($\kappa R/D = 10$, middle panel), the reaction probability is
expectedly reduced, as more frequent failed reaction attempts give
more chances for the molecule to escape to infinity.  This effect is
further enhanced at even smaller reactivity $\kappa R/D = 1$ (right
panel).  In this regime, there is almost no distinction between weakly
reactive targets and the remaining inert surface.

\section{Discussion}
\label{sec:discussion}

We developed a general mathematical description of
diffusion-controlled reactions on catalytic surfaces with
heterogeneous reactivity $\kappa(\s)$.  We showed how the propagator
of the diffusion equation with Robin boundary condition can be
expressed in terms of the Dirichlet propagator for homogeneous
perfectly reactive surface.  The latter involves much simpler and more
studied Dirichlet boundary condition and thus describes exclusively
the first-passage events to the boundary that are independent of the
surface reactivity.  In other words, the diffusive exploration of the
bulk is disentangled from the chemical kinetics on the boundary.  As a
consequence, the Dirichlet propagator needs to be computed only once
for a given geometric configuration, offering a powerful theoretical
and numerical tool for investigating the effects of heterogeneous
surface reactivity.

Numerical or eventually analytical inversion of the Laplace transform
allows one to recover the propagator in time domain.  Moreover, the
Laplace-transformed propagator itself is important as it describes the
steady-state diffusion of molecules which may spontaneously disappear
in the bulk with the rate $p$ \cite{Yuste13,Meerson15,Grebenkov17d}.
Such ``mortal walkers'' may represent radioactive nuclei,
photobleaching fluorophores, molecules in an excited state, metastable
complexes, spermatozoa, and other particles subject to spontaneous
decay, disintegration, ground state recovery, or death.

When the boundary of the domain is bounded, our general representation
yields the spectral decompositions of the propagator and of other
important quantities such as the survival probability, the probability
density of reaction times, the spread harmonic measure, and the
reaction rate.  These decompositions involve the eigenvalues and
eigenfunctions of the Dirichlet-to-Neumann operator, as well as the
associated basis elements of the surface reactivity $\kappa(\s)$.
This spectral description brings new insights onto imperfect
diffusion-controlled reactions and creates a mathematical basis for
formulating and solving optimization and inverse problems on surface
reactivity $\kappa(\s)$ (see, e.g., \cite{Nguyen10,Filoche08}).

We highlight a similarity between the representation of surface
reactivity in the eigenbasis of the Dirichlet-to-Neumann operator and
the representation of the bulk reactivity in the Laplace operator
eigenbasis studied in \cite{Nguyen10}.  Such matrix representations
have proved to be efficient for solving numerically the Bloch-Torrey
equation that describes diffusion magnetic resonance imaging (see
\cite{Grebenkov07,Grebenkov08,Grebenkov16d} and references
therein).  Note that the surface reactivity could also be incorporated
via the Laplacian eigenbasis by introducing an infinitely thin
reactive boundary layer as discussed in
\cite{Grebenkov07a,Grebenkov09}.  However, the eigenbasis of the
Dirichlet-to-Neumann operator acting on the boundary seems to be more
natural for dealing with surface reactivity.  Most importantly, our
spectral description is also valid for exterior problems, for which
the spectrum of the Laplace operator is continuous and thus not
suitable for such representations; in turn, the spectrum of the
Dirichlet-to-Neumann operator on a bounded boundary remains discrete.

We applied the spectral approach to an important example of a
spherical surface, for which both the Dirichlet propagator and the
eigenbasis of the Dirichlet-to-Neumann operator are known explicitly.
In this case, the Robin propagator and the related quantities (such as
the probability density of reaction times) are obtained in a
semi-analytical form, in which the dependence on the starting and
arrival points is fully explicit, whereas the coefficients need to be
computed by truncating and inverting an explicitly known matrix.
However, the proposed approach is not limited to the spherical
boundary.  For instance, the case of a hyperplane was partly studied
in \cite{Sapoval05,Grebenkov19}; apart from straightforward extensions
to disks and cylinders, one can consider more complicated catalytic
surfaces formed by multiple non-overlapping spheres, for which the
Dirichlet propagator in the steady-state regime was recently
investigated in \cite{Grebenkov19b}.  In general, the eigenbasis of
the Dirichlet-to-Neumann operator $\M_p$ can be constructed
numerically; since $\M_p$ is independent of the surface reactivity,
this construction has to be performed only once for a given catalytic
surface.

As mentioned earlier, most former studies focused on the mixed
Dirichlet-Neumann boundary value problem describing perfectly reactive
targets on an otherwise inert boundary.  In spite of its
oversimplified character from the chemical point of view, this problem
may look simpler from the mathematical point of view.  For instance,
as both Dirichlet and Neumann boundary conditions are conformally
invariant, conformal mapping results in a universal integral
representation of the mean first-passage time for planar domains
\cite{Grebenkov16c}.  In addition, the technique of dual series is
more developed for this case \cite{Sneddon,Duffy}.  At the same time,
mixed Dirichlet-Neumann condition is the most problematic from the
perspective of the present work.  Even though the Dirichlet boundary
condition can be formally implemented by setting $\kappa(\s) = \kappa$
on the target and then letting $\kappa$ go to infinity, an infinitely
large jump of reactivity at the border of the target requires
elaborate asymptotic analysis.  In fact, this limit is in general
highly nontrivial because the unbounded Dirichlet-to-Neumann operator
$\M_p$ cannot be neglected as compared to the bounded operator $\K$
(representing the reactivity) even as $\kappa\to\infty$.  This
situation resembles the asymptotic analysis of the Schr\"odinger
operator $-h^2 \Delta + V$ in the semi-classical limit $h\to 0$, where
$V$ is a bounded potential.  While the application of asymptotic
techniques from spectral theory and quantum mechanics to our setting
presents an interesting mathematical perspective for future research,
our spectral approach is not well suited for studying mixed
Dirichlet-Neumann boundary value problems.  

Similarly, in the narrow escape limit (when targets are very small), a
large number of eigenfunctions of the Dirichlet-to-Neumann operator is
needed to accurately represent the multiplication operator $\K$ by a
truncated matrix $\KK$, making numerical computations time-consuming.
More generally, when the shape of the boundary is rather complex or
not smooth enough (e.g., containing corners or cusps), the computation
of the Dirichlet-to-Neumann eigenfunctions becomes difficult, whereas
a large number of eigenfunctions may be needed to project even a
smooth surface reactivity.  In other words, when the surface
reactivity has a substantial projection on a large number of
eigenfunctions, the ``effective'' dimensionality of the matrix $\KK$
can be large, making the proposed spectral approach less efficient
from the numerical point of view.  Nevertheless, the present approach
can still be advantageous for exterior problems, which are
particularly difficult to deal with by other numerical techniques.  In
this light, the present approach does not substitute conventional
techniques but aims to complement them by addressing imperfect
diffusion-controlled reactions on catalytic surfaces with finite
continuously varying heterogeneous reactivity.

\appendix
\section{Alternative representation based on the fundamental solution}
\label{sec:ADirichlet}

In this Appendix, we describe an alternative scheme for representing
the propagator in terms of the fundamental solution of the modified
Helmholtz equation.

\subsection{Dirichlet propagator and the Dirichlet-to-Neumann operator}

The Laplace-transformed Dirichlet propagator $\tilde{G}_0(\x,p|\x_0)$
and the Dirichlet-to-Neumann operator $\M_p$ are closely related.  On
one hand, the action of $\M_p$ onto a given function can be expressed
via Eq. (\ref{eq:Mp_j0}) in terms of the propagator
$\tilde{G}_0(\x,p|\x_0)$ by solving the corresponding Dirichlet
boundary value problem.  On the other hand, the Dirichlet propagator
can be constructed explicitly from the Dirichlet-to-Neumann operator.
For this purpose, one can first represent the propagator as
\begin{equation} \label{eq:G0_Phi}
\tilde{G}_0(\x,p|\x_0) = \G(\x,p|\x_0) + \hat{g}_0(\x,p|\x_0),
\end{equation}
where 
\begin{equation}
\G(\x,p|\x_0) = \frac{\exp(-|\x-\x_0| \sqrt{p/D})}{4\pi D |\x-\x_0|}
\end{equation}
is the fundamental solution of the modified Helmholtz equation in
three dimensions, 
\begin{equation}  \label{eq:Phip}
(p-D\Delta)\G(\x,p|\x_0) = \delta(\x-\x_0), 
\end{equation}
whereas $\hat{g}_0(\x,p|\x_0)$ is the regular part of the propagator
satisfying, for any fixed $\x_0 \in \overline{\Omega}$,
\begin{subequations}
\begin{eqnarray}
(p - D\Delta) \hat{g}_0(\x,p|\x_0) &=& 0 \quad (\x \in \Omega) ,\\  \label{eq:g0_BC}
\hat{g}_0(\x,p|\x_0) + \G(\x,p|\x_0) &=& 0 \quad (\x \in \pa).
\end{eqnarray}
\end{subequations}
Here we use hat instead of tilde in order to distinguish the involved
quantities from those in Sec. \ref{sec:spectral}.

The above problem can be solved in a standard way by using the
Dirichlet propagator $\tilde{G}_0(\x,p|\x_0)$:
\begin{equation*}
\hat{g}_0(\x,p|\x_0) = \int\limits_\pa d\s \, \hat{g}_0(\s,p|\x_0) 
\biggl(-D \frac{\partial \tilde{G}_0(\x',p|\x)}{\partial \n_{\x'}}\biggr)_{\x'=\s} .
\end{equation*}
Using the boundary condition (\ref{eq:g0_BC}) and substituting the
representation (\ref{eq:G0_Phi}), one gets
\begin{eqnarray}  \label{eq:g0_Phi2}
\hat{g}_0(\x,p|\x_0) &=& \int\limits_\pa d\s \, \bigl(- \G(\s,p|\x_0)\bigr) \\   \nonumber
&\times& \biggl(\j(\s,p|\x) + [D \M_p \G(\cdot,p|\x)](\s)\biggr),
\end{eqnarray}
where
\begin{equation}
\j(\s,p|\x_0) = -D \left. \biggl(\frac{\partial \G(\x,p|\x_0)}{\partial \n_{\x}} \biggr)\right|_{\x=\s} \quad (\s\in\pa)
\end{equation}
is also a fully explicit function, and we used the
Dirichlet-to-Neumann operator $\M_p$, acting on $\G(\s',p|\x)$ as a
function of a boundary point $\s'$, to represent the normal derivative
of $\hat{g}_0(\x',p|\x)$.  Combining Eqs. (\ref{eq:G0_Phi},
\ref{eq:g0_Phi2}), we get the representation of the Dirichlet
propagator in terms of the Dirichlet-to-Neumann operator $\M_p$ and
fully explicit functions $\G$ and $\j$.

\subsection{General Robin boundary value problem}

Similarly, for a given function $\tilde{f}(\s,p)$ on the boundary
$\pa$, the solution $\tilde{u}(\x,p)$ of a general Robin boundary
value problem
\begin{subequations}
\begin{eqnarray}  \label{eq:Aauxil1}
(p - D\Delta) \tilde{u} &=& 0 \quad  (\x\in\Omega), \\   \label{eq:Aauxil2}
\biggl(D \frac{\partial}{\partial \n} + \kappa(\x)\biggr) \tilde{u} &=& \tilde{f} \quad (\x\in\pa) ,
\end{eqnarray}
\end{subequations}
can be obtained by multiplying Eqs. (\ref{eq:Phip}, \ref{eq:Aauxil1})
by $\tilde{u}(\x,p)$ and $\G(\x,p|\x_0)$ respectively, subtracting
them, integrating over $\x\in\Omega$, and applying the Green's
formula:
\begin{eqnarray}  \nonumber
\tilde{u}(\x_0,p) &=& \int\limits_{\pa} d\s \biggl(D\G(\s,p|\x_0) \left. \frac{\partial \tilde{u}(\x,p)}{\partial \n_{\x}}\right|_{\x=\s}\\
\label{eq:Aauxil5}
&+& \tilde{u}(\s,p) \j(\s,p|\x_0) \biggr).
\end{eqnarray}
This is a standard representation of a solution of the modified
Helmholtz equation in terms of the surface integral with the potential
$\G(\x,p|\x_0)$ and its normal derivative $\j(\s,p|\x_0)$.  Here,
$\tilde{u}(\x_0,p)$ in a bulk point $\x_0\in\Omega$ is determined by
its values and its normal derivative on the boundary.  In turn, the
Robin boundary condition (\ref{eq:Aauxil2}) can be expressed in terms
of the Dirichlet-to-Neumann operator $\M_p$ and the operator $\K$ of
multiplication by $\kappa(\x)/D$ as $\tilde{u}(\s,p) =
\frac{1}{D}[(\M_p + \K)^{-1} \tilde{f}](\s)$, from which
Eq. (\ref{eq:Aauxil5}) yields
\begin{align}  \nonumber
\tilde{u}(\x_0,p) &= \int\limits_{\pa} d\s \biggl(\G(\s,p|\x_0) \,[\M_p (\M_p + \K)^{-1} \tilde{f}](\s) \\
&+ \frac{1}{D}[(\M_p + \K)^{-1} \tilde{f}](\s) \, \j(\s,p|\x_0)\biggr). 
\end{align}
Since the operator $\M_p$ is self-adjoint, this solution can also be
written as
\begin{align}  \nonumber
\tilde{u}(\x_0,p) &= \frac{1}{D} \int\limits_{\pa} d\s \biggl(\j(\s,p|\x_0) + [D\M_p \G(\cdot,p|\x_0)](\s)\biggr) \\
&\times [(\M_p + \K)^{-1} \tilde{f}](\s) .
\end{align}
If the boundary $\pa$ is bounded, the spectrum of $\M_p$ is discrete,
and this solution can be written as a spectral decomposition:
\begin{eqnarray}    \nonumber
\tilde{u}(\x_0,p) &=& \frac{1}{D} \sum\limits_{n,n'=0}^\infty \hat{V}_{n}^{(p)}(\x_0) \bigl[(\MM + \KK)^{-1}\bigr]_{n,n'}   \\
&\times& \int\limits_{\pa} d\s \, [v_{n'}^{(p)}(\s)]^*\,  \tilde{f}(\s,p),
\end{eqnarray}
where 
\begin{equation}  \label{eq:AhatV}
\hat{V}_n^{(p)}(\x_0) = \int\limits_{\pa} d\s \, v_n^{(p)}(\s) \biggl(\j(\s,p|\x_0) + D \mu_n^{(p)} \G(\s,p|\x_0)  \biggr),
\end{equation}
and the matrices $\MM$ and $\KK$ are defined in Eq. (\ref{eq:MK_def}).
In particular, the Laplace-transformed propagator
$\tilde{G}(\x,p|\x_0)$ for the Robin boundary value problem
(\ref{eq:tildeG_eq0}) can be written as
\begin{eqnarray} \nonumber
&& \tilde{G}(\x,p|\x_0) = \G(\x,p|\x_0) + \frac{1}{D} \sum\limits_{n,n'=0}^\infty \hat{V}_{n}^{(p)}(\x_0)  \\  \label{eq:Gtilde2A}
&& \times \bigl[(\MM + \KK)^{-1}\bigr]_{n,n'} \, [\hat{U}_{n'}^{(p)}(\x)]^* ,
\end{eqnarray}
where
\begin{equation} \label{eq:AhatU}
\hat{U}_{n}^{(p)}(\x) = \int\limits_{\pa} d\s \, v_{n}^{(p)}(\s) \biggl(\j(\s,p|\x) - \kappa(\s) \G(\s,p|\x)\biggr)  \,.
\end{equation}
In contrast to Eq. (\ref{eq:Gtilde2}), this representation is based on
the explicitly known fundamental solution $\G(\x,p|\x_0)$ and does not
involve the Dirichlet propagator $\tilde{G}_0(\x,p|\x_0)$.  As a
consequence, all the deduced spectral decompositions rely uniquely on
the eigenbasis of the Dirichlet-to-Neumann operator.  While the
representations (\ref{eq:Gtilde2}) and (\ref{eq:Gtilde2A}) are
equivalent and complementary to each other, we keep using the former
one due to its simpler form and clearer probabilistic interpretation.

\section{Technical derivations}

\subsection{Reaction probability}
\label{sec:reacprob}

The reaction probability can be obtained by integrating the
probability density $H(t|\x_0)$ of reaction times over $t$ from $0$ to
infinity, giving $\tilde{H}(0|\x_0)$.  For any bounded domain, a
diffusing molecule cannot avoid the reaction event so that
$\tilde{H}(0|\x_0) = 1$, ensuring the correct normalization of the
probability density $H(t|\x_0)$.  This property can be checked
directly from our spectral representation (\ref{eq:Htilde}).  Setting
$p = 0$ yields the reaction probability
\begin{eqnarray}  \nonumber
\tilde{H}(0|\x_0) &=& \sum\limits_{n,n'=0}^\infty V_n^{(0)}(\x_0) \bigl[(\MM+\KK)^{-1}\KK \bigr]^{(p=0)}_{n,n'} \\
\label{eq:Aauxil6}
&\times& \int\limits_\pa d\s \, [v_{n'}^{(0)}(\s)]^*  .
\end{eqnarray}

For any bounded domain, the Laplace equation $\Delta u = 0$ with
$u_{|\pa} = 1$ on the boundary has the constant solution, $u \equiv
1$, so that a constant function $1$ on the boundary is an
eigenfunction of the Dirichlet-to-Neumann operator, $v_0^{(0)}(\s) =
|\pa|^{-1/2}$, corresponding to $\mu_0^{(0)} = 0$.  As a consequence,
the second sum over $n'$ in Eq. (\ref{eq:Aauxil6}) vanishes due to the
orthogonality of eigenfunctions, yielding
\begin{equation}
\tilde{H}(0|\x_0) = \sum\limits_{n=0}^\infty V_n^{(0)}(\x_0) \bigl[(\MM+\KK)^{-1}\KK \bigr]^{(p=0)}_{n,0} |\pa|^{1/2} .
\end{equation}
Rewriting $(\MM+\KK)^{-1}\KK$ as $\I - (\MM+\KK)^{-1} \MM$ and using
the diagonal structure of $\MM$, one gets
\begin{eqnarray}  \label{eq:H0_int} 
\tilde{H}(0|\x_0) &=& V_0^{(0)}(\x_0) |\pa|^{1/2} - |\pa|^{1/2}\\  \nonumber
&\times& \sum\limits_{n=0}^\infty V_n^{(0)}(\x_0) \bigl[(\MM+\KK)^{-1}\KK \bigr]^{(p=0)}_{n,0} \underbrace{\mu_0^{(0)}}_{=0}  \\  \nonumber
&=& \int\limits_\pa d\s \, \tilde{j}_0(\s,0|\x_0) = 1,
\end{eqnarray}
where the last integral reflects the normalization of the harmonic
measure density $\tilde{j}_0(\s,0|\x_0)$ for a bounded domain.

For an unbounded domain, a nonzero constant cannot be a solution of
the Laplace equation $\Delta u = 0$ with $u_{|\pa} = 1$ due to the
regularity condition $u(\x) \to 0$ as $|\x|\to\infty$.  The function
$\M_0 1$ is thus not zero, and the smallest eigenvalue $\mu_0^{(0)}$
is strictly positive.  As a consequence, the second term in
Eq. (\ref{eq:H0_int}) does not vanish, while the first term is not
equal to $1$.  In other words, the reaction probability
$\tilde{H}(0|\x_0)$ is in general less than $1$ due to the possibility
for a molecule to escape at infinity.  In this case, $H(t|\x_0)$
should be renormalized by $\tilde{H}(0|\x_0)$ to get the conditional
probability density of reaction times.

\subsection{Laplace-transformed reaction rate}
\label{sec:AVn}

We briefly discuss how Eq. (\ref{eq:Ktilde}) for the
Laplace-transformed reaction rate $\tilde{J}(p)$ can be further
simplified when the initial concentration is uniform: $c_0(\x_0) =
c_0$.

Integrating Eq. (\ref{eq:tildeG_eq}) for the propagator
$\tilde{G}_0(\x,p|\x_0)$ over $\x\in \Omega$ yields
\begin{equation}
p \int\limits_\Omega d\x_0 \, \tilde{G}_0(\x,p|\x_0) = 1 + D \int\limits_{\pa} d\x_0
\frac{\partial \tilde{G}_0(\x,p|\x_0)}{\partial \n_{\x_0}} \,,
\end{equation}
where we exchanged $\x$ and $\x_0$ due to the symmetry of the
propagator.  Applying the normal derivative at a boundary point $\x =
\s \in \pa$ and multiplying by $-D$, we get 
\begin{equation}
\int\limits_\Omega d\x_0 \, \tilde{j}_0(\s,p|\x_0) = \frac{D}{p} \int\limits_{\pa} d\s_0
\left. \left(
\frac{\partial \tilde{j}_0(\s,p|\x_0)}{\partial \n_{\x_0}} \right) \right|_{\x_0=\s_0} \,.
\end{equation}
Multiplying this relation by a function $f(\s)$ and integrating over
$\s\in\pa$, we have
\begin{eqnarray*}
&& \int\limits_\Omega d\x_0 \, \int\limits_\pa d\s\, \tilde{j}_0(\s,p|\x_0) \, f(\s) \\
&& = \frac{D}{p} \int\limits_{\pa} d\s_0 \left. \left(
\frac{\partial}{\partial \n_{\x_0}} \int\limits_{\pa} d\s \,\tilde{j}_0(\s,p|\x_0) \, f(\s) \right) \right|_{\x_0=\s_0} \\
&& = \frac{D}{p} \int\limits_{\pa} d\s_0 \, [\M_p f](\s_0) ,
\end{eqnarray*}
where the order of integrals was exchanged.  As it is satisfied for
any $f(\s)$, we conclude that
\begin{equation}
[\M_p 1](\s) = \frac{p}{D} \int\limits_\Omega d\x_0 \, \tilde{j}_0(\s,p|\x_0).
\end{equation}
Setting $f(\s) = v_n^{(p)}(\s)$, we also deduce
\begin{equation} \label{eq:integal_Vn}
\int\limits_\Omega d\x_0 \, V_n^{(p)}(\x_0) = \frac{D}{p} \mu_n^{(p)} \int\limits_{\pa} d\s \, v_n^{(p)}(\s) .
\end{equation}
This expression allows us to compute the integral in
Eq. (\ref{eq:Ktilde}), yielding
\begin{eqnarray}  \label{eq:Ktilde2}
\tilde{J}(p) &=& \frac{c_0 D}{p} \sum\limits_{n,n'=0}^\infty \left(\int\limits_\pa d\s \, v_n^{(p)}(\s) \right) \\  \nonumber
&\times& \bigl[\MM (\MM + \KK)^{-1} \KK\bigr]_{n,n'} \left(\int\limits_\pa d\s \, [v_{n'}^{(p)}(\s)]^* \right) .
\end{eqnarray}

\section{Diffusion inside a ball}
\label{sec:Ainterior}

Solutions of Dirichlet boundary value problems for the modified
Helmholtz equation in a ball and the related operators are well known.
For the sake of clarify and completeness, we summarize the main
``ingredients'' involved in our spectral decompositions.

To determine the eigenbasis of the Dirichlet-to-Neumann operator in a
ball of radius $R$, $\Omega = \{ \x\in\R^3~:~ |\x| < R\}$, one simply
notes that a general solution of the modified Helmholtz equation $(p -
D\Delta) \tilde{u} = 0$ can be written in spherical coordinates
$(r,\theta,\phi)$ as
\begin{equation}
\tilde{u}(\x,p) = \sum\limits_{n=0}^{\infty} \sum\limits_{m=-n}^m a_{mn} \, i_n(r\sqrt{p/D})\, Y_{mn}(\theta,\phi) ,
\end{equation}
where $a_{mn}$ are unknown coefficients, 
\begin{equation}
i_n(z) = \sqrt{\pi/2} \, \frac{I_{n+1/2}(z)}{\sqrt{z}}
\end{equation}
are the modified spherical Bessel functions of
the first kind, and 
\begin{equation}
Y_{mn}(\theta,\phi) = c_{nm} \, P_n^m(\cos\theta) \, e^{im\phi} 
\end{equation}
are the spherical harmonics, with $P_n^m(x)$ being the associated
Legendre functions and $c_{nm}$ the normalization coefficients:
\begin{equation}  \label{eq:cnm}
c_{nm} = \sqrt{\frac{2n+1}{4\pi} \, \frac{(n-m)!}{(n+m)!}} \,.
\end{equation}
As the normal derivative of $\tilde{u}$ on the boundary involves only
the radial coordinate and does not affect $Y_{mn}(\theta,\phi)$, the
eigenvalues and eigenfunctions of the Dirichlet-to-Neumann operator
are
\begin{subequations}  \label{eq:mu_v_int}
\begin{eqnarray}  \label{eq:mu_nm_sphereA}
\mu_{nm}^{(p)} &=& \sqrt{p/D} \, \frac{i'_n(R\sqrt{p/D})}{i_n(R\sqrt{p/D})} \,, \\  \label{eq:vnm_sphereA}
v_{nm}(\theta,\phi) &=& \frac{1}{R} \, Y_{mn}(\theta,\phi) ,
\end{eqnarray}
\end{subequations}
where prime denotes the derivative with respect to the argument.  As
stated in the main text, the double index $nm$ is employed to
enumerate the eigenfunctions as well as the elements of the matrices
$\MM$ and $\KK$.  Note that the eigenvalues that determine the matrix
$\MM$ in Eq. (\ref{eq:MK_def}), do not depend on the index $m$ and
thus are of multiplicity $2n+1$.  In turn, the eigenfunctions $v_{nm}$
do not depend on the parameter $p$ that will simplify further
expressions.  The explicit computation of the matrix $\KK$ from
Eq. (\ref{eq:KK_sphere}) is discussed in Appendix \ref{sec:K}.

For a ball, the Dirichlet propagator is known explicitly
\begin{eqnarray} \label{eq:GDir_int}
G_0(\x,t|\x_0) &=& \frac{1}{2\pi R^3} \sum\limits_{n=0}^\infty (2n+1) P_n\biggl(\frac{(\x \cdot \x_0)}{|\x|\, |\x_0|}\biggr)  \\ \nonumber
&\times& \sum\limits_{k=0}^\infty \frac{e^{-Dt\alpha_{nk}^2/R^2}}{[j'_n(\alpha_{nk})]^2}
 j_n(\alpha_{nk} r/R)  \, j_n(\alpha_{nk} r_0/R) ,
\end{eqnarray}
where $\alpha_{nk}$ are the positive zeros (enumerated by the index
$k=0,1,2,\ldots$) of the spherical Bessel functions $j_n(z)$ of the
first kind, and we used the addition theorem for spherical harmonics
to evaluate the sum over the index $m$:
\begin{eqnarray}  \label{eq:addition}
P_n\biggl(\frac{(\x \cdot \x_0)}{|\x|\, |\x_0|}\biggr) &=& 
4\pi \sum\limits_{m=-n}^n \frac{Y_{mn}(\theta_0,\phi_0) Y_{mn}^*(\theta,\phi)}{2n+1} \\  \nonumber
&=&  \frac{4\pi R^2}{2n+1} \sum\limits_{m=-n}^n v_{nm}(\theta_0,\phi_0) \, v_{nm}^*(\theta,\phi),
\end{eqnarray}
where $P_n(z)$ are the Legendre polynomials.  The Laplace transform of
Eq. (\ref{eq:GDir_int}) reads
\begin{eqnarray} \nonumber
\tilde{G}_0(\x,p|\x_0) &=& \frac{1}{2\pi R^3} \sum\limits_{n=0}^\infty (2n+1) P_n\biggl(\frac{(\x \cdot \x_0)}{|\x|\, |\x_0|}\biggr) \\  
\label{eq:G0_int}
&\times& \sum\limits_{k=0}^\infty \frac{j_n(\alpha_{nk} r/R)  \, j_n(\alpha_{nk} r_0/R)}{(D\alpha_{nk}^2/R^2 + p)[j'_n(\alpha_{nk})]^2} \,.
\end{eqnarray}
As $\frac{(\x \cdot \x_0)}{|\x|\, |\x_0|}$ is the cosine of the angle
between the vectors $\x$ and $\x_0$, it does not depend on the radial
coordinates $r$ and $r_0$.  We get thus
\begin{eqnarray} \nonumber
\tilde{j}_0(\s,p|\x_0) &=& - \frac{1}{2\pi R^2} \sum\limits_{n=0}^\infty (2n+1) P_n\biggl(\frac{(\s \cdot \x_0)}{|\s|\, |\x_0|}\biggr) \\ 
\label{eq:j0_int}
&\times& \sum\limits_{k=0}^\infty \frac{\alpha_{nk} \, j_n(\alpha_{nk} r_0/R)}{(\alpha_{nk}^2 + pR^2/D) j'_n(\alpha_{nk})} \,,
\end{eqnarray}
from which
\begin{eqnarray} \nonumber 
V_{nm}^{(p)}(\x_0) &=& - 2v_{nm}(\theta_0,\phi_0) \sum\limits_{k=0}^\infty   
\frac{\alpha_{nk} j_n(\alpha_{nk} r_0/R)}{(\alpha_{nk}^2 + pR^2/D) j'_n(\alpha_{nk})} \\  \label{eq:Vnm_int}
&=& v_{nm}(\theta_0,\phi_0) \, \frac{i_n(r_0\sqrt{p/D})}{i_n(R\sqrt{p/D})} \,,
\end{eqnarray}
where we used the summation formula over zeros $\alpha_{nk}$ (see
Eq. (S9) from Table 3 of Ref. \cite{Grebenkov19c}).  Expressions
(\ref{eq:G0_int}, \ref{eq:j0_int}, \ref{eq:Vnm_int}) determine the
Laplace-transformed propagator $\tilde{G}(\x,p|\x_0)$ of the Robin
boundary value problem in the semi-analytical form (\ref{eq:Gtilde2}),
in which the dependence on points $\x_0$ and $\x$ is fully explicit,
whereas the computation of the coefficients involves a numerical
inversion of the matrix $\MM + \KK$.
Similarly, we deduce semi-analytical expressions for the
Laplace-transformed probability density of reaction times and the
spread harmonic measure presented in the main text.  For instance, as
the eigenfunctions $v_{nm}$ are orthogonal to $v_{00}(\theta,\phi) =
1/(\sqrt{4\pi}R)$, the sum in Eq. (\ref{eq:hn}) is reduced to a
single term, yielding Eq. (\ref{eq:hnm}).

We also compute the mean reaction time $\tilde{S}(0|\x_0)$ by
evaluating the limit $p\to 0$ of the Laplace-transformed survival
probability from Eq. (\ref{eq:Stilde}).  In this limit, one gets
\begin{equation}
\mu_{nm}^{(0)} = n/R, \quad V_{nm}^{(0)}(\x_0) = v_{nm}(\theta_0,\phi_0) (r_0/R)^n, 
\end{equation}
so that
\begin{eqnarray}  \nonumber
&& \tilde{S}(0|\x_0) = \sqrt{4\pi} \sum\limits_{n=0}^\infty \sum\limits_{m=-n}^n Y_{mn}(\theta_0,\phi_0) (r_0/R)^n  \\  \label{eq:Tmean}
&& \times \biggl(\frac{R^2 - r_0^2}{4D(n+3/2)} h_{mn}^{(0)} - \biggl(\frac{h_{mn}^{(p)}}{dp}\biggr)_{p=0} \biggr).
\end{eqnarray}
The last term can be evaluated explicitly as
\begin{equation}
\biggl(\frac{h_{mn}^{(p)}}{dp}\biggr)_{p=0} = - \bigl[(\MM_0 + \KK)^{-1} \MM_1 (\MM_0 + \KK)^{-1} \KK\bigr]_{nm,00} ,
\end{equation}
where $\MM_0$ and $\MM_1$ are diagonal matrices obtained by expanding
the elements of $\MM$ into powers $p$: $[\MM_0]_{nm,n'm'} =
\delta_{n,n'} \delta_{m,m'} n/R$ and $[\MM_1]_{nm,n'm'} =
\delta_{n,n'} \delta_{m,m'} \frac{R/D}{2n+3}$.

For a homogeneous reactivity, $\kappa(\theta,\phi) = \kappa$,
Eqs. (\ref{eq:KK_hom}, \ref{eq:Htilde_sphere_main}) imply
\begin{equation} \label{eq:Htilde_sphere_uni}
\tilde{H}_{\rm hom}(p|\x_0) = \frac{\kappa\, i_0(r_0\sqrt{p/D})}{\sqrt{pD} \, i'_0(R\sqrt{p/D}) + \kappa \, i_0(R\sqrt{p/D})} 
\end{equation}
(since $i_0(z) = \sinh(z)/z$, one can further simplify this
expression).  In turn, Eq. (\ref{eq:Ktilde_sphere}) gives
\begin{equation}
\tilde{J}_{\rm hom}(p) = 4\pi D R c_0 \biggl(\frac{i_0(R\sqrt{p/D})}{R\sqrt{p/D} \, i_1(R\sqrt{p/D})} + \frac{D}{\kappa R}\biggr)^{-1} \,,
\end{equation}
and its inverse Laplace transform yields an infinite sum of
exponentially decaying functions with the rates determined by the
poles of this expression (see \cite{Carslaw,Crank,Redner}).  Finally,
Eq. (\ref{eq:Tmean}) yields the classical result
\begin{equation}  \label{eq:Tmean_hom}
\tilde{S}_{\rm hom}(0|\x_0) = \frac{R^2 - r_0^2}{6D} + \frac{R}{3\kappa} \,.
\end{equation}

\subsection*{Numerical validation}

\begin{figure}
\begin{center}
\includegraphics[width=88mm]{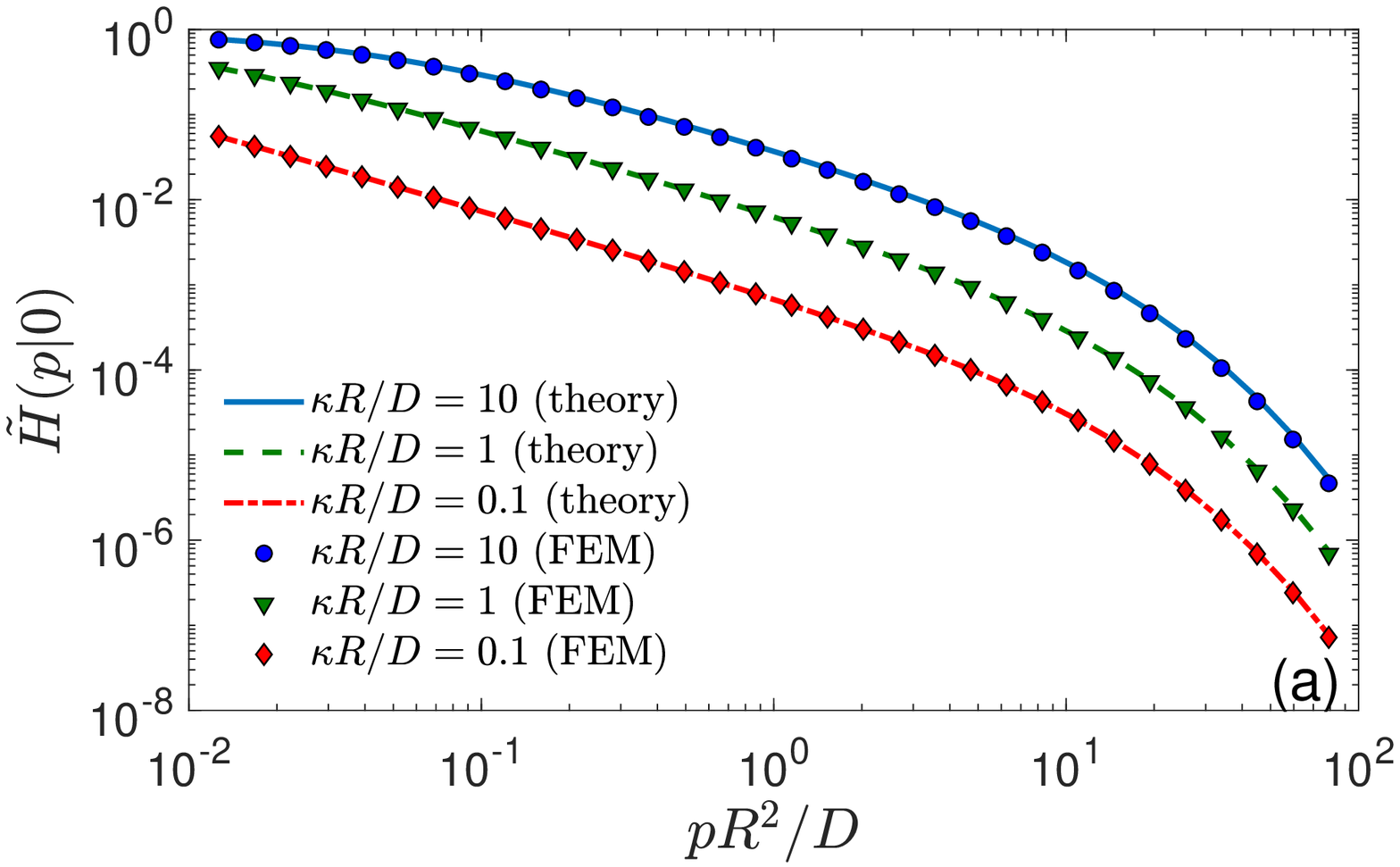} 
\includegraphics[width=88mm]{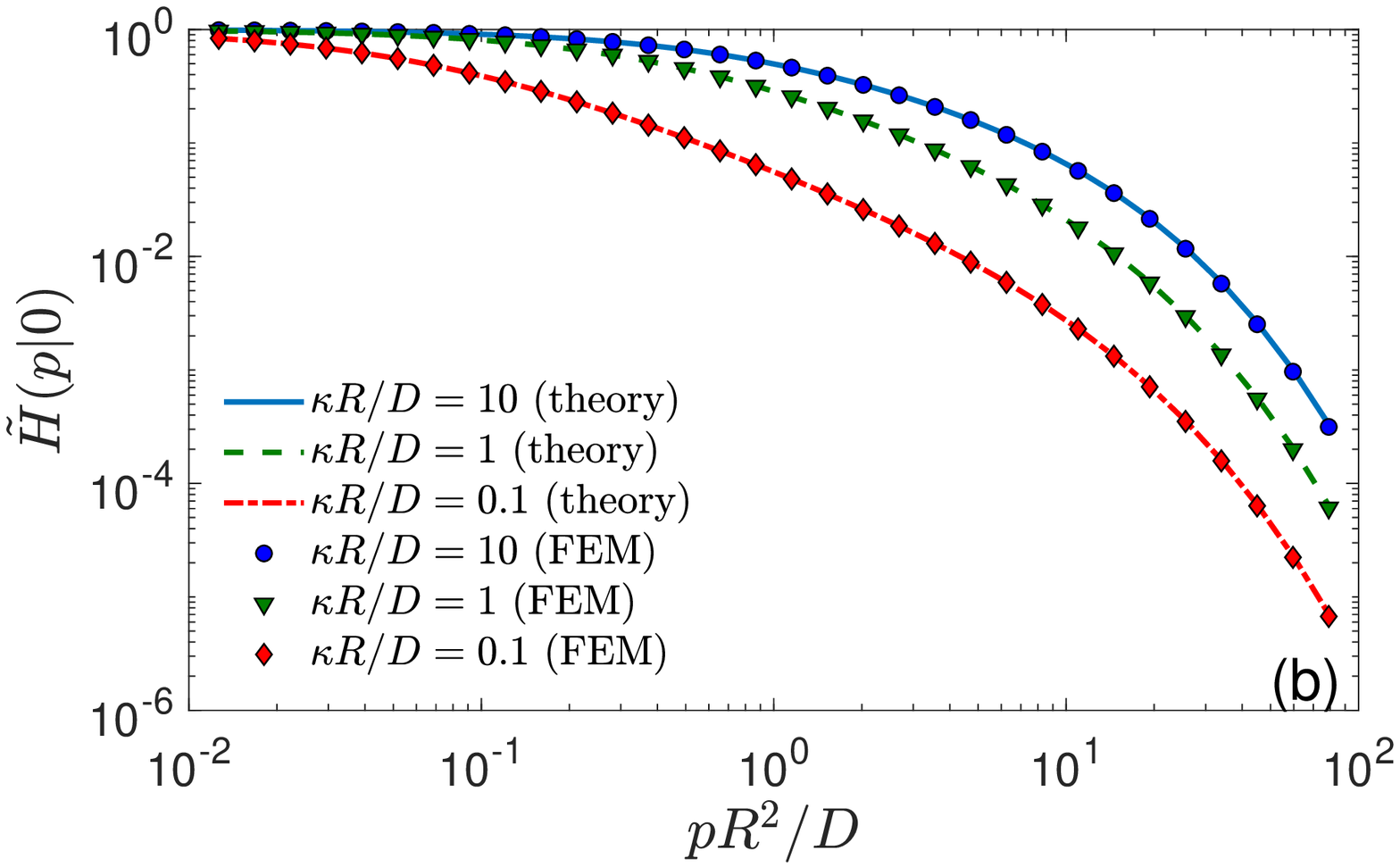} 
\end{center}
\caption{
Laplace-transformed probability density $\tilde{H}(p|0)$ of reaction
times on a partially reactive circular target of reactivity $\kappa$
and angular size $\ve$, located on the inert spherical surface of
radius $R$, for a molecule started from the origin, with $\ve = 0.1$
{\bf (a)} and $\ve = 1$ {\bf (b)}.  Lines show the semi-analytical
solution (\ref{eq:Hp0_int}), in which $h_{00}^{(p)}$ was found from
Eq. (\ref{eq:hnm}) with the matrices $\MM$ and $\KK$ truncated at
$\nmax = 20$.  Symbols present a FEM numerical solution with the
maximal mesh size of $0.01$.}
\label{fig:Hp_FEM}
\end{figure}

To illustrate the quality of our semi-analytical solution, we look at
the Laplace-transformed probability density $\tilde{H}(p|\x_0)$ that
satisfies the boundary value problem
\begin{eqnarray*}
(p-D\Delta_{\x_0}) \tilde{H}(p|\x_0) &=& 0 \quad (\x_0\in\Omega), \\
\biggl(D \frac{\partial}{\partial \n_{\x_0}} + \kappa(\x_0)\biggr) \tilde{H}(p|\x_0) &=& \kappa(\x_0) \quad (\x_0\in\pa),
\end{eqnarray*}
with $\Delta_{\x_0}$ acting on $\x_0$.  We set $\kappa(\theta,\phi) =
\kappa \, \Theta(\ve - \theta)$ to describe a single partially
reactive circular target of angular size $\ve$ and reactivity
$\kappa$, located at the North pole (here $\Theta(z)$ is the Heaviside
function).  The axial symmetry of this geometric setting allows one to
reduce the original three-dimensional problem to a two-dimensional one
on the rectangle $[0,R] \times [0,\pi]$ in the coordinates
$(r,\theta)$.  We solve this problem by using a finite element method
implemented in Matlab PDE toolbox.  The computational domain was
meshed with the constraint on the largest mesh size to be $0.01$.  For
the sake of simplicity, we fix the starting point at the origin, in
which case Eq. (\ref{eq:Htilde_sphere_main}) is reduced to
\begin{equation} \label{eq:Hp0_int}
\tilde{H}(p|0) = h_{00}^{(p)} \, \frac{R\sqrt{p/D}}{\sinh(R\sqrt{p/D})} \,,
\end{equation}
where $h_{00}^{(p)}$ is given by Eq. (\ref{eq:hnm}) and computed with
the matrices $\KK$ and $\MM$ truncated to the size $\nmax = 20$ and
constructed from Eqs.  (\ref{eq:K_single}, \ref{eq:MK_def}).  Figure
\ref{fig:Hp_FEM} shows an excellent agreement between this
semi-analytical form and the FEM solution for both a small target of
angular size $\ve = 0.1$ (with the surface fraction $\sigma =
(1-\cos\ve)/2 \approx 0.0025$) and a large target of angular size $\ve
= 1$ (with $\sigma \approx 0.23$), and different reactivities.

\section{Computation of the matrix $\KK$}
\label{sec:K}

The key element of the spectral approach is the possibility to
disentangle first-passage diffusive steps from the heterogeneous
reactivity which is incorporated via the matrix $\KK$.  In this
Appendix, we compute this matrix for several most common settings on
the spherical boundary.

\subsection{General setting}
\label{sec:K_general}

In general, the reactivity $\kappa(\theta,\phi)$ can be expanded over
the complete basis of spherical harmonics,
\begin{equation}
\kappa(\theta,\phi) = \sum\limits_{n=0}^\infty \sum\limits_{m=-n}^n \kappa_{nm} Y_{mn}(\theta,\phi),
\end{equation}
with coefficients $\kappa_{nm}$.  When this expansion can be truncated
at a low order $n^*$, one can compute the elements of the matrix $\KK$
explicitly, without numerical quadrature in Eq. (\ref{eq:KK_sphere}),
by using the following identity
\begin{eqnarray}  \nonumber
&& \int\limits_0^\pi d\theta \sin\theta \int\limits_0^{2\pi} d\phi \, Y_{m_1n_1}(\theta,\phi) \, Y_{m_2n_2}(\theta,\phi) \, Y_{m_3n_3}(\theta,\phi) \\
\nonumber
&& = \sqrt{\frac{(2n_1+1)(2n_2+1)(2n_3+1)}{4\pi}} \\   \label{eq:YYY}
&& \times \left(\begin{array}{c c c} n_1 & n_2 & n_3 \\ m_1 & m_2 & m_3 \\ \end{array}\right)
\left(\begin{array}{c c c} n_1 & n_2 & n_3 \\ 0 & 0 & 0 \\ \end{array}\right),
\end{eqnarray}
where $\left(\begin{array}{c c c} n_1 & n_2 & n_3 \\ m_1 & m_2 & m_3
\\ \end{array}\right)$ is the Wigner 3j symbol.
We note that the truncation order $\nmax$ should significantly exceed
$n^*$ to ensure accurate computations.

\subsection{Axially symmetric problems}

When the reactivity is axially symmetric, $\kappa(\theta,\phi) =
\kappa(\theta)$, the integral over $\phi$ in Eq. (\ref{eq:KK_sphere})
yields $2\pi \delta_{m,m'}$, and the matrix $\KK$ has a block
structure.  If in addition one is interested in axially symmetric
quantities (e.g., $\tilde{H}(p|\x_0)$ which does not depend on
$\phi_0$ due to the axial symmetry), it is sufficient to construct a
reduced version of the matrix $\KK$ by eliminating repeated lines and
rows and keeping only the elements with $m = m' = 0$:
\begin{eqnarray}  \label{eq:KK_axial}
\KK_{n0,n'0} &=& \sqrt{(n+1/2)(n'+1/2)} \\ \nonumber
&\times& \int\limits_0^\pi d\theta \, \sin\theta \, \frac{\kappa(\theta)}{D}
\, P_n(\cos\theta) \, P_{n'}(\cos\theta) .
\end{eqnarray}
From the numerical point of view, this drastically speeds up
computations because the size of the matrix $\KK$, truncated to the
order $\nmax$, becomes $(\nmax+1) \times (\nmax+1)$ instead of
$(\nmax+1)^2 \times (\nmax+1)^2$ in the general setting.
Semi-analytical expressions also become simpler, e.g.,
Eqs. (\ref{eq:Htilde_sphere_main}) and (\ref{eq:Htilde_sphere_ext})
read respectively
\begin{equation} \label{eq:Htilde_sphereA}
\tilde{H}(p|\x_0) = \sum\limits_{n=0}^\infty \sqrt{2n+1} \, h_{n0}^{(p)} \, \frac{i_n(r_0\sqrt{p/D})}{i_n(R\sqrt{p/D})} \, P_n(\cos\theta_0) 
\end{equation}
and
\begin{equation} \label{eq:Htilde_sphereA_ext}
\tilde{H}(p|\x_0) = \sum\limits_{n=0}^\infty \sqrt{2n+1} \, h_{n0}^{(p)} \, \frac{k_n(r_0\sqrt{p/D})}{k_n(R\sqrt{p/D})} \, P_n(\cos\theta_0) ,
\end{equation}
with $h_{n0}^{(p)}$ given by Eq. (\ref{eq:hnm}).  

If an expansion of the reactivity $\kappa(\theta)$ over the complete
basis of Legendre polynomials is known,
\begin{equation}  \label{eq:kappa_Pn}
\kappa(\theta) = \sum\limits_{n=0}^\infty \kappa_n \, P_n(\cos\theta),
\end{equation}
then the elements of the matrix $\KK$ can be computed by using the
identity
\begin{equation}
\int\limits_{-1}^1 dx \, P_{n_1}(x) \, P_{n_2}(x) \, P_{n}(x) = 2 \left(\begin{array}{ccc} n_1 & n_2 & n \\  0 & 0 & 0 \\ \end{array}\right)^2 ,
\end{equation}
which follows from Eq. (\ref{eq:YYY}).  As the selection rule for
Wigner 3j-symbols requires that $|n_1 - n_2| \leq n \leq n_1 + n_2$,
the truncation of the expansion (\ref{eq:kappa_Pn}) at the order $n^*$
implies that the matrix $\KK$ has at most $n^*$ subdiagonals above and
below the main diagonal that simplifies the construction of this
matrix.
One advantage of the representation (\ref{eq:kappa_Pn}) is that the
average reactivity is equal to $\kappa_0$ and is independent of
$\kappa_n$ with $n \geq 1$ due to the orthogonality of Legendre
polynomials.

We emphasize however that the above simplified construction is not
sufficient for computing the Laplace-transformed propagator
$\tilde{G}(\x,p|\x_0)$ which is not axially symmetric.  In fact,
Eq. (\ref{eq:Gtilde2}) involves the coefficients $[(\MM +
\KK)^{-1}]_{nm,n'm'}$, whose computation requires all the elements
$\KK_{nm,n'm'}$ even for axially symmetric reactivity, and it is not
reducible to that with the elements $\KK_{n0,n'0}$.  In this case, the
general scheme from Sec. \ref{sec:K_general} should be used.

\subsection{Single circular target}

To model a single circular partially reactive target of angular size
$\ve$ at the North pole (with the remaining inert boundary), one sets
\begin{equation}
\kappa(\theta,\phi) = \kappa\, \Theta(\ve - \theta) ,
\end{equation}
so that Eq. (\ref{eq:KK_axial}) yields
\begin{equation}
\KK_{n0,n'0} = \frac{\kappa}{D} \sqrt{(n+1/2)(n'+1/2)} \int\limits_{\cos\ve}^1 dx \, P_n(x) \, P_{n'}(x) .
\end{equation}
To compute explicitly the matrix $\KK$, one can use the
Adams-Neumann's product formula (see \cite{Al-Salam57}):
\begin{equation}
P_n(x) P_{n'}(x) = \sum\limits_{k=0}^{\min\{n,n'\}} B_{nn'}^k  \, P_{n+n'-2k}(x),
\end{equation}
where 
\begin{equation}
B_{nn'}^k = \frac{A_k A_{n-k} A_{n'-k}}{A_{n+n'-k}} \, \frac{2n+2n'-4k+1}{2n+2n'-2k+1} \,,
\end{equation}
with
$A_k = 
\frac{\Gamma(k+1/2)}{\sqrt{\pi} \Gamma(k+1)}$ (with $A_0 = 1$).  We
get thus
\begin{eqnarray}  \label{eq:K_single}
&& \KK_{n0,n'0} = \frac{\kappa}{D} \sqrt{(n+1/2)(n'+1/2)} \times \\  \nonumber
&& \sum\limits_{k=0}^{\min\{n,n'\}} \hspace*{-2mm} B_{nn'}^k \, \frac{P_{n+n'-2k-1}(\cos\ve) - P_{n+n'-2k+1}(\cos\ve)}{2(n+n'-2k)+1} \,,
\end{eqnarray} 
where we used the identity for $n \geq 0$
\begin{equation}  \label{eq:Pn_int}
\int\limits_a^b dx \, P_n(x) = \frac{P_{n+1}(b) - P_{n-1}(b) - P_{n+1}(a) + P_{n-1}(a)}{2n+1} 
\end{equation}
(with the convention $P_{-1}(x) = 1$).

An explicit formula for $\KK$ is also easily deducible for multiple
latitudinal stripes.  The domain is still axially symmetric and one
just needs to sum up contributions from each stripe, relying on the
explicit integral of $P_n(x)$ in Eq. (\ref{eq:Pn_int}).

\subsection{Multiple targets of circular shape}
\label{sec:multiple}

The matrix $\KK$ can also be computed explicitly for multiple
partially reactive non-overlapping targets of circular shape.  In
fact, the additivity of the integral in Eq. (\ref{eq:KK_sphere})
implies that contributions for all targets are just summed up.  We
consider thus the contribution of the $i$-th target $\Gamma_i$ of
angle $\ve_i$, reactivity $\kappa_i$, and the angular coordinates
$(\theta_i,\phi_i)$ for its center.  It is convenient to apply the
rotational addition theorem for spherical harmonics to rotate the
coordinate system \cite{Steinborn73}: 
\begin{equation}
Y_{mn}(\theta',\phi') = \sum\limits_{m'=-n}^n [D_{mm'}^{n}(\phi_i, \theta_i, \phi_i)]^* \, Y_{m'n}(\theta,\phi),
\end{equation}
where $D_{mm'}^n(\alpha, \beta, \gamma)$ is the Wigner D-matrix
describing the rotation by Euler angles $(\alpha,\beta,\gamma)$.  As a
consequence, the $i$-th contribution to the matrix $\KK$ reads
\begin{eqnarray*} 
&& \KK_{n_1m_1,n_2m_2}^{(i)} = \frac{\kappa_i}{DR^2}  \int\limits_{\Gamma_i} d\s' \,  
Y_{m_1n_1}^*(\theta',\phi') \, Y_{m_2n_2}(\theta',\phi')  \\  
&&= \frac{\kappa_i}{DR^2} \sum\limits_{m_1'=-n_1}^{n_1} D_{m_1m'_1}^{n_1}(\phi_i,\theta_i,\phi_i) \sum\limits_{m_2'=-n_2}^{n_2}  \\
&& \times  [D_{m_2m'_2}^{n_2}(\phi_i,\theta_i,\phi_i)]^{*}
 \int\limits_{\Gamma_0} d\s \, Y_{m_1'n_1}^*(\theta,\phi) Y_{m_2'n_2}(\theta,\phi)  ,
\end{eqnarray*}
where $\Gamma_0$ is the $i$-th target rotated to be centered around
the North pole.  To proceed, one can express the product of two
spherical harmonics as 
\begin{eqnarray} \label{eq:YY0}
&& Y_{m'_1n_1}(\theta,\phi) \, Y_{m'_2n_2}(\theta,\phi) \\  \nonumber
&&=  \sum\limits_{n=|n_1-n_2|}^{n_1+n_2} B_{m'_1n_1m'_2n_2}^{n} Y_{(m'_1+m'_2)n}(\theta,\phi)
\end{eqnarray}  
(which follows from Eq. (\ref{eq:YYY})), where
\begin{eqnarray} 
&& B_{m'_1n_1m'_2n_2}^{n} = \sqrt{\frac{(2n+1)(2n_1+1)(2n_2+1)}{4\pi}}  \\  \nonumber
&& \times (-1)^{m'_1+m'_2} \left(\begin{array}{c c c} n_1 & n_2 & n \\ m'_1 & m'_2 & -m'_1-m'_2 \\ \end{array}\right)
\left(\begin{array}{c c c} n_1 & n_2 & n \\ 0 & 0 & 0 \\ \end{array}\right),
\end{eqnarray}
with $\biggl(\begin{array}{c c c} n_1 & n_2 & n \\ m_1 & m_2 & m \\
\end{array}\biggr)$ being again the Wigner 3-j symbols \cite{Brink},
and we employ the convention that $Y_{mn}(\theta,\phi) \equiv 0$ if
$|m| > n$.  Using the identity
\begin{equation*}
Y_{mn}^*(\theta,\phi) = (-1)^m  Y_{(-m)n}(\theta,\phi),
\end{equation*}
the above formula yields
\begin{eqnarray} \label{eq:YY}
&& Y_{m'_1n_1}^*(\theta,\phi) \, Y_{m'_2n_2}(\theta,\phi) \\  \nonumber
&& = (-1)^{m'_2} \sum\limits_{n=|n_1-n_2|}^{n_1+n_2} B_{(-m'_1)n_1m'_2n_2}^{n} \, Y_{(m'_2-m'_1)n}(\theta,\phi),
\end{eqnarray} 
from which
\begin{widetext}
\begin{eqnarray}   \nonumber
&& \KK_{n_1m_1,n_2m_2}^{(i)} = \sqrt{\pi} \, \frac{\kappa_i}{D} \sum\limits_{n=|n_1-n_2|}^{n_1+n_2} 
\, \frac{P_{n-1}(\cos\ve_i) - P_{n+1}(\cos\ve_i)}{\sqrt{2n+1}} \\
&& \times \sum\limits_{m=-\min\{n_1,n_2\}}^{\min\{n_1,n_2\}} (-1)^m D_{m_1m}^{n_1}(\phi_i,\theta_i,\phi_i)
[D_{m_2m}^{n_2}(\phi_i,\theta_i,\phi_i)]^* \, B_{(-m)n_1mn_2}^{n}  \,,
\end{eqnarray}
\end{widetext}
where the integral over $\phi$ yielded $2\pi \delta_{m'_1,m'_2}$ that
removed one sum, while the integral of $P_n(x)$ was evaluated from
Eq. (\ref{eq:Pn_int}).  We get therefore a fully explicit expression
for the contribution of the $i$-th target to the matrix $\KK$.  One
can thus compute the Laplace-transformed propagator in the
semi-analytical form, as for a single target.

\section{Diffusion outside a ball}
\label{sec:Aexterior}

For diffusion in the unbounded domain $\Omega = \{\x\in\R^3 ~:~ |\x| >
R\}$ outside a ball of radius $R$, the eigenfunctions of the
Dirichlet-to-Neumann operator are still given by
Eq. (\ref{eq:vnm_sphere}), whereas the eigenvalues are
\begin{equation}  \label{eq:mu_sphere_extA}
\mu_{nm}^{(p)} = - \sqrt{p/D} \, \frac{k'_n(R\sqrt{p/D})}{k_n(R\sqrt{p/D})} \,, 
\end{equation}
where $k_n(z)$ are the modified spherical Bessel functions of the
second kind:
\begin{equation}
k_n(z) = \sqrt{2/\pi}\, \frac{K_{n+1/2}(z)}{\sqrt{z}} \,.  
\end{equation}
As for the interior problem, the eigenvalue $\mu_{nm}^{(p)}$ does not
depend on $m$ and has thus the multiplicity $2n+1$.  Note that
$\mu_{nm}^{(p)}$ are just polynomials of $R\sqrt{p/D}$, e.g.,
$\mu_{00}^{(p)} = (1 + R\sqrt{p/D})/R$.

The Laplace-transformed Dirichlet propagator is known:
\begin{eqnarray*}
&& \tilde{G}_0(\x,p|\x_0) = \frac{e^{-\sqrt{p/D}|\x - \x_0|}}{4\pi D|\x - \x_0|}- \frac{\sqrt{p/D}}{4\pi D} \sum\limits_{n=0}^\infty
(2n+1)  \\ 
&& \times P_n\biggl(\frac{(\x\cdot \x_0)}{|\x|\, |\x_0|}\biggr) \frac{i_n(R\sqrt{p/D})}{k_n(R\sqrt{p/D})} k_n(r\sqrt{p/D}) k_n(r_0\sqrt{p/D}).
\end{eqnarray*}
Note that the fundamental solution (the first term) also admits the
decomposition:
\begin{eqnarray}\nonumber
\frac{e^{-\sqrt{p/D}|\x - \x_0|}}{4\pi |\x - \x_0|} &=& \frac{\sqrt{p/D}}{4\pi} \sum\limits_{n=0}^\infty 
(2n+1) P_n\biggl(\frac{(\x\cdot \x_0)}{|\x| \, |\x_0|}\biggr) \\  
&\times& k_n(r_0\sqrt{p/D}) \, i_n(r\sqrt{p/D}) 
\end{eqnarray}
for $r < r_0$ (and $r_0$ is exchanged with $r$ for $r > r_0$), where
we applied the addition theorem (\ref{eq:addition}) for spherical
harmonics.  One gets then
\begin{eqnarray}  \label{eq:G0_sphere_ext}
&& \hspace*{-3mm} 
\tilde{G}_0(\x,p|\x_0) = \frac{\sqrt{p/D}}{4\pi D} \sum\limits_{n=0}^\infty
(2n+1)P_n\biggl(\frac{(\x\cdot \x_0)}{|\x|\, |\x_0|}\biggr) \times  \\  \nonumber
&& \hspace*{-3mm}  
k_n(r_0\sqrt{p/D}) \biggl(i_n(r\sqrt{p/D}) - k_n(r\sqrt{p/D}) \frac{i_n(R\sqrt{p/D})}{k_n(R\sqrt{p/D})} \biggr)
\end{eqnarray}
for $r < r_0$.  In particular, one deduces
\begin{equation}  \label{eq:j0_ext}
\tilde{j}_0(\s,p|\x_0) = \sum\limits_{n=0}^\infty
\frac{2n+1}{4\pi R^2} P_n\biggl(\frac{(\s\cdot \x_0)}{|\s|\, |\x_0|}\biggr) \frac{k_n(r_0\sqrt{p/D})}{k_n(R\sqrt{p/D})} \,,
\end{equation}
where we used the Wronskian $i'_n(z) k_n(z) - k'_n(z) i_n(z) = 1/z^2$.
We compute then
\begin{equation}  \label{eq:Vnm_sphere_ext}
V_{nm}^{(p)} = v_{nm}(\theta_0,\phi_0) \frac{k_n(r_0\sqrt{p/D})}{k_n(R\sqrt{p/D})} \,.
\end{equation}
According to our spectral decomposition (\ref{eq:Gtilde2}),
Eqs. (\ref{eq:G0_sphere_ext}, \ref{eq:Vnm_sphere_ext}), together with
the matrices $\MM$ and $\KK$, fully determine the Laplace-transformed
propagator $\tilde{G}(\x,p|\x_0)$.  Similarly, we deduce the
Laplace-transformed probability density of reaction times and the
reaction time presented in the main text.

In the limit $p\to 0$, one has 
\begin{equation}  \label{eq:mu_p0}
\mu_{nm}^{(0)} = (n+1)/R, \quad V_{nm}^{(0)}(\x_0) = v_{nm}(\theta_0,\phi_0) (R/r_0)^{n+1} \,,
\end{equation}
from which
\begin{equation} \label{eq:H0_ext}
\tilde{H}(0|\x_0) 
= \sqrt{4\pi} \sum\limits_{n=0}^\infty \sum\limits_{m=-n}^n  h_{nm}^{(0)} \, (R/r_0)^{n+1} Y_{mn}(\theta_0,\phi_0) 
\end{equation}
is the probability of reaction on the ball, and $h_{nm}^{(p)}$ are
defined by Eq. (\ref{eq:hnm}).  In contrast to bounded domains, for
which this probability was equal to $1$ (see Eq. (\ref{eq:H0_int})),
the transient character of Brownian motion in three dimensions makes
this probability less than $1$.  Rewriting Eq. (\ref{eq:hnm}) as
\begin{equation}
h_{nm}^{(0)} = \delta_{n,0} \delta_{m,0} - \frac{1}{R} \bigl[(\MM_{p=0} + \KK)^{-1}\bigr]_{nm,00} ,
\end{equation}
one can split $\tilde{H}(0|\x_0)$ into two terms, in which the first
term $R/r_0$ is the hitting probability to a perfectly reactive ball,
while the second term accounts for partial heterogeneous reactivity.
The limit $p\to 0$ also determines the long-time behavior of
Eq. (\ref{eq:Ktilde_sphere}) that yields Eq. (\ref{eq:k_steady}) for
the steady-state reaction rate $J(\infty)$.

For homogeneous reactivity, $\kappa(\s) = \kappa$, $\MM + \KK$ is a
diagonal matrix and thus only the term with $m=n=0$ survives in
Eq. (\ref{eq:H0_ext}), yielding the classical result for the hitting
probability of a homogeneous partially reactive ball:
\begin{equation}
\tilde{H}_{\rm hom}(0|\x_0) = \frac{R}{r_0} \, \frac{1}{1 + D/(\kappa R)} \,.
\end{equation}
More generally, Eq. (\ref{eq:Htilde_sphere_ext}) yields
\begin{equation} \label{eq:Htilde_sphere_ext_uni}
\tilde{H}_{\rm hom}(p|\x_0) = \frac{R}{r_0(1 + \frac{D}{\kappa R})} \, 
\frac{e^{-(r_0-R)\sqrt{p/D}}}{1 + \frac{R\sqrt{p/D}}{1 + \kappa R/D}} \,,
\end{equation}
where we used $k_0(z) = e^{-z}/z$.  The Laplace inversion recovers the
result by Collins and Kimball \cite{Collins49}
\begin{eqnarray}   \label{eq:Ht_Rinf}
&& H_{\rm hom}(t|\x_0) = \frac{\kappa}{r_0}\exp\biggl(-\frac{(r_0-R)^2}{4Dt}\biggr)
\biggl\{\frac{R}{\sqrt{\pi Dt}}  \\  \nonumber
&& - \biggl(1+\frac{\kappa R}{D}\biggr)\erfcx\biggl(\frac{r_0-R}{\sqrt{4Dt}}+\biggl(1+\frac{\kappa R}{D}\biggr)
\frac{\sqrt{Dt}}{R}\biggr) \biggr\},
\end{eqnarray}
where $\erfcx(x)=e^{x^2} \erfc(x)$ is the scaled complementary error
function (see also discussion in \cite{Grebenkov18c}).  In the limit
$\kappa\to\infty$, this expression reduces to
\begin{equation}
\label{eq:Ht_Rinf_kinf}
H_{\rm hom}(t|\x_0)=\frac{R}{r_0} \, \frac{r_0-R}{\sqrt{4\pi Dt^3}} \, \exp\biggl(-\frac{(r_0-R)^2}{4Dt}\biggr) .
\end{equation}
Finally, Eq. (\ref{eq:Ktilde_sphere}) gives after simplifications
\begin{equation}   
\tilde{J}_{\rm hom}(p) = \frac{4\pi D R c_0}{1 + \frac{D}{\kappa R}} \biggl(\frac{1}{p}+ \frac{\kappa R/D}{p +  
(1 + \frac{\kappa R}{D}) \sqrt{pD}/R}\biggr),
\end{equation}
from which one retrieves in time domain the reaction rate derived by
Collins and Kimball \cite{Collins49} 
\begin{subequations}
\begin{eqnarray}   \label{eq:kt_Collins}
\hspace*{-5mm}
\frac{J_{\rm hom}(t)}{J_{\rm hom}(\infty)} &=& 1 + 
\frac{\kappa R}{D} \erfcx\biggl(\biggl(1 + \frac{\kappa R}{D}\biggr) \frac{\sqrt{Dt}}{R}\biggr) ,  \\  \label{eq:kt_Collins2}
\hspace*{-5mm}
J_{\rm hom}(\infty) &=& \frac{4\pi D R c_0}{1 + \frac{D}{\kappa R}} \,.
\end{eqnarray}
\end{subequations}
In the short-time limit $t\to 0$, the reaction rate approaches a
constant, $J_{\rm hom}(0) = 4\pi \kappa R^2 c_0$, which corresponds to
reaction-limited kinetics (note that $J_{\rm hom}(0) > J_{\rm
hom}(\infty)$).  In the limit $\kappa\to\infty$, one retrieves the
Smoluchowski result:
\begin{equation}
J_{\rm hom}(t) = 4\pi D R c_0 \biggl(1 + \frac{\sqrt{R}}{\sqrt{\pi Dt}}\biggr) \,.
\end{equation}

We note that the above analysis can be easily extended to the
case when the spherical target of radius $R$ is surrounded by an outer
reflecting concentric sphere of radius $R_o$.  The eigenfunctions of
the Dirichlet-to-Neumann operator remain unchanged, whereas the
eigenvalues become
\begin{widetext}
\begin{equation}
\mu_n^{(p)} = -\sqrt{p/D}\, \frac{k'_n(R_o\sqrt{p/D})\, i'_n(R \sqrt{p/D}) - i'_n(R_o\sqrt{p/D})\, k'_n(R \sqrt{p/D})}
{k'_n(R_o\sqrt{p/D})\, i_n(R \sqrt{p/D}) - i'_n(R_o\sqrt{p/D})\, k_n(R \sqrt{p/D})} \,.
\end{equation}
\end{widetext}
In the limit $R_o\to \infty$, one retrieves
Eq. (\ref{eq:mu_sphere_extA}) for the exterior of a ball.  As $p\to
0$, one also gets
\begin{equation}
\mu_n^{(0)} = \frac{n+1}{R} \, \frac{1-(R/R_o)^{2n+1}}{1 + (1+1/n) (R/R_o)^{2n+1}} \,.
\end{equation}
As a consequence, one can easily extend the former results to this
setting.


\end{document}